\documentclass[12pt]{article}
\pdfoutput=1
\usepackage{amssymb,amsthm,amsmath,amsfonts}
\usepackage{graphicx}
\usepackage{microtype}
\usepackage{xcolor}
\usepackage[pdftex,bookmarks=true,colorlinks,linkcolor=red,urlcolor=blue,citecolor=blue]{hyperref}
\usepackage{cite}
\usepackage{subcaption}
\usepackage{float}
\usepackage{bm}
\usepackage{tocloft}
\usepackage{placeins}
\usepackage{needspace}
\def\baselinestretch{1.1}
\makeatletter\@addtoreset{equation}{section}\makeatother

\newcommand{\preprint}[1]{\begin{table}[t]\begin{flushright}{#1}\end{flushright}\end{table}}
\renewcommand{\title}[1]{\vbox{\center\LARGE{#1}}\vspace{5mm}}
\renewcommand{\author}[1]{\vbox{\center#1}\vspace{5mm}}
\newcommand{\address}[1]{\vbox{\center\em#1}}

\begin{document}

\hypersetup{pageanchor=false}
\begin{titlepage}
\vspace{.5cm}
\preprint{}
\date{\today}
\begin{center}
\vskip 0.2cm
\title{\boldmath From Critical Zero Modes to Interior Dynamics in Holographic Superconductors}
\vskip 0.5cm
{Yan Liu$^{\,a}$}\footnote{Email: {\tt yanliu@ytu.edu.cn}},
{Hong-Da Lyu$^{\,b}$}\footnote{Email: {\tt hongdalyu@buaa.edu.cn}}

\address{${}^{a}$ Department of Physics, Yantai University, 30 Qingquan Road, Yantai 264005, China}
\address{${}^{b}$Key Laboratory of Particle Physics and Particle Irradiation (MOE),\\Institute of Frontier and Interdisciplinary Science,\\Shandong University, Qingdao, Shandong 266237, China}
\end{center}
\vskip 1.0cm

\abstract{
We study how the critical behavior of holographic superconductors
controls the interior geometry of charged black holes. Although the exterior condensate vanishes at criticality, the critical zero mode develops logarithmic oscillations near the Reissner--Nordstr\"om Cauchy horizon, rendering the interior critical limit nonuniform. We show that the transmitted critical-mode amplitude, together with local
Cauchy-horizon quantities, fixes both the leading Einstein--Rosen collapse $\Delta z_{\rm ER}$ width and
the phase $\Theta_J$ in the subsequent Josephson regime. If the condensate obeys
$t\equiv1-T/T_c\propto\epsilon^{2r}$,
then
$\Delta z_{\rm ER}\propto t^{1/r}$
and
$\Theta_J\propto t^{-1/r}$. This Josephson phase is transferred to the first Kasner epoch, producing
oscillatory Kasner exponents and an accumulation law
$t_n\sim n^{-r}$ for successive zeros of the scalar velocity. The critical zero mode fixes the leading collapse and phase coefficients,
which are quantitatively confirmed by fully nonlinear solutions in both
the ordinary and tricritical cases. At low temperature, the interior instead depends on the infrared completion of the zero-temperature geometry. Thus near-critical scaling is governed by critical-mode transmission, whereas the low-temperature behavior is controlled by the infrared endpoint.}

\vfill
\end{titlepage}

\hypersetup{pageanchor=true}
\begingroup
\hypersetup{linkcolor=black}
\renewcommand{\baselinestretch}{1}\fontsize{10}{12}\selectfont
\setlength{\parskip}{0pt}
\setlength{\cftbeforesecskip}{5pt}
\setlength{\cftbeforesubsecskip}{1pt}
\tableofcontents
\endgroup
\newpage

\section{Introduction}
\label{sec:intro}

Holographic duality provides a gravitational description of phase
transitions in strongly coupled quantum systems. Thermodynamic
quantities, order parameters, and critical exponents are usually
extracted from the asymptotic region or the black-hole exterior
~\cite{Zaanen:2015book,Hartnoll:2016matter}. The same bulk solutions,
however, also determine the geometry behind the event horizon. This
raises a natural question: to what extent does boundary critical
behavior determine the black-hole interior?

Scalar deformations provide a useful setting to explore this question.
Relevant deformations of a thermal CFT can drive the bulk geometry
from the AdS boundary through the horizon toward a Kasner
singularity~\cite{Frenkel:2020Kasner}. The resulting Kasner exponents
can also reflect the boundary phase structure: they may develop
nonanalytic behavior across holographic phase
transitions~\cite{Liu:2021hap}, while their evolution near topological phase
transitions has been studied in holographic Weyl and nodal-line
semimetals~\cite{Gao:2023semimetals}.
More recently, the scaling of the condensate near ordinary and
multicritical bifurcations was related perturbatively to the first
Kasner epoch by following the critical scalar mode into the
interior~\cite{Gao:2026tck}. Related critical Kasner scaling has
also been found near the onset of spontaneous scalarization in
asymptotically flat charged black
holes~\cite{Li:2025critical,Sun:2026scaling}.

The situation is qualitatively different when the normal phase contains
a Cauchy horizon. Perturbations are strongly amplified near an inner
horizon, as familiar from mass inflation and strong cosmic
censorship~\cite{Poisson:1989mass,Dafermos:2003CH}, while quantum fields
can develop singular stress tensors there
~\cite{ShahbaziMoghaddam:2024CH}. Dynamical scalar collapse also
exhibits critical thresholds for singularity formation in charged
black holes and for inner horizon contraction in regular Hayward
geometries~\cite{Shao:2025critical,Shao:2025Hayward}.
In charged AdS black holes, even weak
scalar hair can destroy the regular inner horizon and replace it by a
rapid Einstein-Rosen (ER) collapse, followed by Josephson oscillations
in the charged condensate case and eventually a spacelike Kasner
singularity~\cite{Hartnoll:2020rwq,Hartnoll:2020fhc,Cai:2020wrp}.
Related behavior has since been found in a broad class of hairy black
holes, including rotating, anisotropic, and modified-gravity
examples~\cite{Gao:2024rotating,Dias:2021interior,Sword:2021pfm,Mansoori:2021interior,
Liu:2022helical,Carballo:2024metals,Arean:2024interiors}, while nonlinear
matter interactions can generate Kasner inversions and repeated
transitions~\cite{Zhang:2024multiband,Zhao:2025nonlinear,
An:2022Mtheory,Cai:2021vector,Cai:2024vector,Henneaux:2022Kasner,
Cai:2023igv,Gao:2026analytic3D}.

For neutral scalar bifurcations, Ref.~\cite{Gao:2026CH} showed that the
critical zero mode regular at event horizon develops a logarithmic branch
near the reference Cauchy horizon. The coefficient of this branch fixes
the perturbation entering the nonlinear collapse layer, allowing the
leading ER-collapse scale and, when a kinetic overlap exists, the first
Kasner behavior to be determined directly from the normalized critical
mode. This provides a quantitative transmission map from the exterior
critical mode to the interior in the neutral case.

For holographic superconductors, the condensing scalar is charged and
the post-collapse interior contains an additional Josephson regime.
The local nonlinear dynamics of this regime was analyzed in
Refs.~\cite{Hartnoll:2020fhc}, where the ER collapse
was matched to oscillatory Bessel dynamics and subsequently to Kasner
evolution. To relate this sequence to the superconducting transition,
we match the critical scalar zero mode of the reference RN geometry to
the nonlinear ER collapse and trace its consequences for the Josephson
and first Kasner regimes.

We find that the transmitted critical zero mode fixes the leading
near-critical interior scales. If the exterior condensate amplitude is
$\epsilon$, the ER-collapse width scales as
$\Delta z_{\rm ER}\propto\epsilon^2$, while the accumulated Josephson
phase grows as $\Theta_J\propto\epsilon^{-2}$. Their coefficients are
determined by the transmission amplitude and local quantities at the
reference RN Cauchy horizon. Combining this result with the exterior
critical law
$t\equiv1-T/T_c\propto\epsilon^{2r}$ gives
\begin{equation}
\Delta z_{\rm ER}\propto t^{1/r},
\qquad
\Theta_J\propto t^{-1/r}.
\end{equation}

\begin{figure}[h!]
\centering
\includegraphics[width=\textwidth]{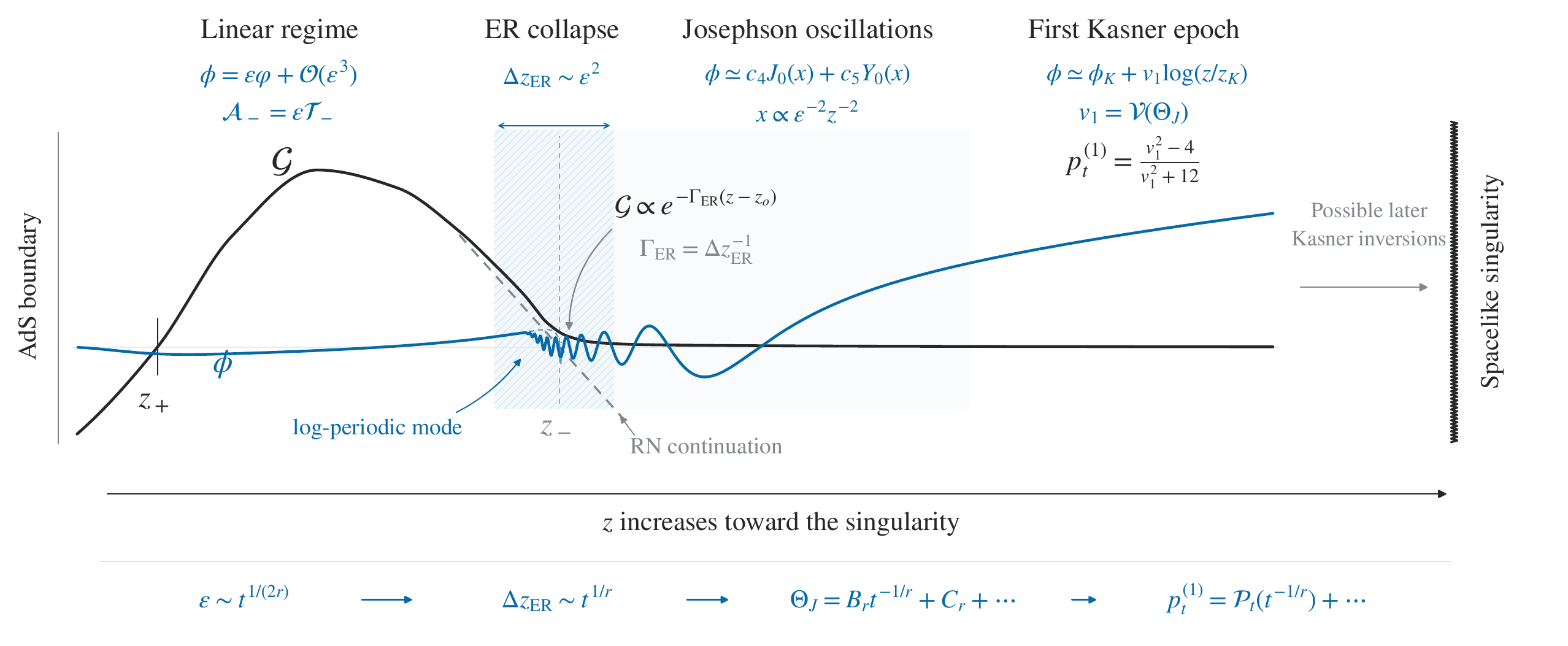}
\caption{\small
Critical zero-mode transmission through the reference Cauchy-horizon
region, the nonlinear ER collapse, and the subsequent Josephson regime.
The incoming Josephson phase is transferred to the first Kasner epoch.
The schematic is not to scale.
}
\label{fig:interior_schematic}
\end{figure}

The Josephson phase is subsequently transferred to the first Kasner
epoch, producing rapidly oscillating Kasner exponents near criticality.
For a simple zero of the transfer function, successive zeros of the
first-Kasner scalar velocity accumulate as
\begin{equation}
t_n\sim n^{-r}.
\end{equation}
Thus an ordinary critical point with $r=1$ gives $t_n\sim n^{-1}$,
whereas the tricritical point considered here has $r=2$ and hence
$t_n\sim n^{-2}$. At tricriticality, the quartic St\"uckelberg
interaction also generates a logarithmic correction to the Josephson
phase without changing its leading $t^{-1/2}$ divergence. These
predictions are confirmed by the fully nonlinear solutions in both the
ordinary and tricritical cases. The near-critical mechanism is summarized schematically in
Fig.~\ref{fig:interior_schematic}.

The opposite limit $T\to0$ probes a different source of interior
structure. We consider three representative nonlinear completions that
share the same critical RN background and linear zero mode but flow to
distinct zero-temperature infrared geometries: a cohesive AdS$_4$
endpoint, a finite-scalar Lifshitz endpoint, and a running-scalar
solution. Their low-temperature interiors are correspondingly different.
The cohesive branch approaches its limiting Kasner exponents by a power
law, the Lifshitz branch approaches finite non-Schwarzschild values with
fractional-power corrections, while the running-scalar branch converges
only inverse logarithmically. Thus the near-critical interior is
controlled by the transmission of the critical zero mode, whereas the
low-temperature behavior is determined by the zero-temperature infrared
completion.

The remainder of the paper is organized as follows.
Sec.~\ref{sec:model} introduces the generalized holographic
superconductor and the black-hole solutions considered below.
Sec.~\ref{sec:multicritical} develops the near-critical perturbation
theory, introduces the representative nonlinear completions, and
analyzes their thermodynamic critical behavior.
Sec.~\ref{sec:interior} derives the transmission of the critical zero
mode into the ER collapse, Josephson regime, and first Kasner epoch.
Sec.~\ref{sec:low_temperature} derives the low-temperature Kasner
scaling associated with the three zero-temperature infrared endpoints.
Sec.~\ref{sec:numerical} compares these analytic predictions with the
full nonlinear solutions.
We conclude in Sec.~\ref{sec:discussion}.

\section{Generalized holographic superconductor}
\label{sec:model}

We consider the four-dimensional generalized holographic
superconductor~\cite{Aprile:2010yb,Hartnoll:2008superconductors}
\begin{align}
S
=
\int d^4x\sqrt{-g}
\left(
R
-\frac14 G(\phi)F^{ab}F_{ab}
-V(\phi)
-\frac12(\nabla\phi)^2
-\frac12J(\phi)A_aA^a
\right)\,,
\label{eq:action}
\end{align}
where $F_{ab}=\partial_aA_b-\partial_bA_a$ is the Maxwell field strength. Eq.~\eqref{eq:action} is the unitary-gauge form of
a generalized St\"uckelberg theory, in which the phase of the charged
order parameter has been absorbed into the gauge field~\cite{Aprile:2010yb}. Throughout this paper, we set $16\pi G_N=L=1$.

The normal phase admits a unit-radius AdS$_4$ vacuum at $\phi=0$, for which
$G(0)=1$, $V(0)=-6$, and $J(0)=0$.  Assuming invariance under $\phi\to-\phi$, we take the couplings to be analytic and even about the normal phase,
\begin{align}
G(\phi)
&=
1+g_2\phi^2+g_4\phi^4+g_6\phi^6+\cdots,
\nonumber\\
V(\phi)
&=
-6-\phi^2+v_4\phi^4+v_6\phi^6+\cdots,
\nonumber\\
J(\phi)
&=
q^2\phi^2+\eta_4\phi^4+\eta_6\phi^6+\cdots ,
\label{eq:couplings}
\end{align}
where the scalar mass is $m^2=-2$.  The quadratic couplings $(g_2,q)$
enter the linear scalar equation, whereas the higher-order coefficients
affect only the nonlinear completion.  We further require a positive Maxwell kinetic coupling,
i.e. $G(\phi)>0$.

We consider homogeneous planar black branes with the ansatz
\begin{align}
ds^2
&=
\frac1{z^2}
\left(
-f(z)e^{-\chi(z)}dt^2
+\frac{dz^2}{f(z)}
+dx^2+dy^2
\right),
\nonumber\\
A
&=
\Phi(z)dt,
\qquad
\phi=\phi(z).
\label{eq:ansatz}
\end{align}
The AdS boundary lies at $z=0$, while increasing $z$ takes the geometry through the event horizon and into the black-hole interior.

For the ansatz~\eqref{eq:ansatz}, the radial equations of motion are
\begin{align}
\chi'
&=
\frac{z}{2}\phi'^2
+
\frac{ze^\chi J(\phi)\Phi^2}{2f^2},
\label{eq:chi_eom}
\\
\left(
\frac{e^{-\chi/2}f}{z^3}
\right)'
&=
\frac{e^{-\chi/2}}{4z^4}
\left(
z^4e^\chi G(\phi)\Phi'^2
+
2V(\phi)
\right)
,
\label{eq:f_eom}
\\
\left(
e^{\chi/2}G(\phi)\Phi'
\right)'
&=
\frac{e^{\chi/2}J(\phi)}{z^2f}\Phi,
\label{eq:maxwell_eom}
\\
\left(
\frac{e^{-\chi/2}f}{z^2}\phi'
\right)'
&=
\frac{e^{-\chi/2}}{z^4}
\left(
V_{,\phi}
-\frac{z^4e^\chi}{2}G_{,\phi}\Phi'^2
-\frac{z^2e^\chi}{2f}J_{,\phi}\Phi^2
\right)
.
\label{eq:scalar_eom}
\end{align}
Here and below, primes denote derivatives with respect to $z$, while
a subscript $,\phi$ denotes differentiation with respect to $\phi$.

We impose source-free standard quantization at the AdS boundary,
\begin{equation}
\phi=\phi_2 z^2+\cdots,
\qquad
\Phi=\mu-\rho z+\cdots,
\qquad
\chi\rightarrow0 ,
\end{equation}
and work in the grand canonical ensemble at fixed chemical potential $\mu=1$.

A regular nonextremal event horizon at $z=z_h$ is characterized by $f(z_h)=
\Phi(z_h)=0$,
with all remaining fields regular there. The Hawking temperature and
entropy density are
\begin{equation}
T
=
-\frac{e^{-\chi(z_h)/2}}{4\pi}f'(z_h),
\qquad
s
=
\frac{4\pi}{z_h^2}.
\label{eq:thermodynamics}
\end{equation}
The corresponding UV and near-horizon expansions are collected in
Appendix~\ref{app:boundary_horizon}. 

\section{Near-critical perturbation theory and thermodynamics}
\label{sec:multicritical}

We construct the superconducting branch perturbatively around the
RN--AdS$_4$ normal phase and introduce the dimensionless radial
coordinate
\begin{equation}
u\equiv\frac{z}{z_h},
\qquad
0\leq u\leq1,
\end{equation}
so that the AdS boundary and the event horizon fixed at $u=0$
and $u=1$, respectively.

\subsection{Perturbative framework}
\label{subsec:sc_expansion}

In standard quantization, a source-free scalar behaves near the
boundary $u=0$ as
\begin{equation}
\phi(u)
=
\epsilon u^2+\cdots,
\label{eq:sc_epsilon_definition}
\end{equation}
We use $\epsilon$ as the expansion parameter. Since $z=z_hu$, the
boundary condensate is
$O=\epsilon/z_h^2$.

The $\phi\rightarrow-\phi$ symmetry implies that the scalar field contains
only odd powers of $\epsilon$, whereas the metric, gauge field, and
horizon position contain only even powers. We therefore expand
\begin{equation}
\begin{split}
\phi
&=
\epsilon\varphi
+\epsilon^3\phi_3
+\epsilon^5\phi_5
+\mathcal O(\epsilon^7),
\\
f
&=
f_0
+\epsilon^2f_2
+\epsilon^4f_4
+\mathcal O(\epsilon^6),
\\
\chi
&=
\epsilon^2\chi_2
+\epsilon^4\chi_4
+\mathcal O(\epsilon^6),
\\
\Phi
&=
\Phi_0
+\epsilon^2\Phi_2
+\epsilon^4\Phi_4
+\mathcal O(\epsilon^6),
\\
z_h
&=
z_c
+\epsilon^2z_2
+\epsilon^4z_4
+\mathcal O(\epsilon^6).
\end{split}
\label{eq:sc_expansions}
\end{equation}
Near the boundary $u=0$, we fix the normalization of the perturbative
expansion by
\begin{equation}
\varphi=u^2+O(u^3),
\qquad
\phi_3=o(u^2),
\qquad
\phi_5=o(u^2).
\label{eq:sc_scalar_normalization}
\end{equation}
The coefficient of the normalizable mode is then entirely encoded in
$\epsilon$..

The boundary conditions at $u=0$ follow from fixed chemical potential
and time normalization,
\begin{equation}
\Phi_0(0)=1,
\qquad
\Phi_{2n}(0)=0,
\qquad
\chi_{2n}(0)=0,
\qquad
f_{2n}(0)=0,
\qquad n\geq1.
\label{eq:sc_UV_order_conditions}
\end{equation}
At the event horizon $u=1$, regularity imposes
\begin{equation}
\Phi_0(1)=0,
\qquad
\Phi_{2n}(1)=0,
\qquad
f_0(1)=0,
\qquad
f_{2n}(1)=0,
\qquad n\geq1,
\label{eq:sc_horizon_order_conditions}
\end{equation}
with all scalar perturbations regular there.

At zeroth order, $\phi_0=0$ and the background solution is the planar
RN--AdS$_4$ black brane,
\begin{equation}
\begin{split}
f_0(u)
&=
1-
\left(
1+\frac{z_c^2}{4}
\right)u^3
+\frac{z_c^2}{4}u^4,
\qquad
\Phi_0(u)
=
1-u,
\qquad
\chi_0=0.
\end{split}
\label{eq:sc_critical_RN}
\end{equation}
The horizon radius $z_c$ is fixed by the scalar zero-mode condition
below. The corresponding critical temperature is
\begin{equation}
T_c
=
\frac{3-z_c^2/4}{4\pi z_c}.
\label{eq:critical_temperature}
\end{equation}

\subsection{Ordinary critical behavior}
\label{subsec:critical_bifurcation}

At first order in $\epsilon$, the scalar perturbation obeys
\begin{equation}
{\cal L}_c\varphi
\equiv
\left(
\frac{f_0}{u^2}\varphi'
\right)'
+
\left(
\frac{2}{u^4}
+g_2z_c^2
+\frac{q^2z_c^2(1-u)^2}{u^2f_0}
\right)\varphi
=
0.
\label{eq:sc_zero_mode}
\end{equation}
Regularity at the event horizon gives
\begin{equation}
\varphi'(1)
=
\frac{2+g_2z_c^2}{3-z_c^2/4}
\varphi(1).
\label{eq:sc_zero_mode_horizon}
\end{equation}
Together with the boundary conditions, Eq.~\eqref{eq:sc_zero_mode}
defines an eigenvalue problem for $z_c$.
We choose the lowest nodeless mode, which corresponds to the highest
critical temperature.

At second order, the critical zero mode backreacts on the metric and gauge
field. The redshift correction satisfies
\begin{equation}
\chi_2'
=
\frac{u}{2}\varphi'^2
+
\frac{q^2z_c^2u(1-u)^2}{2f_0^2}\varphi^2,
\label{eq:sc_chi2}
\end{equation}
while the Maxwell equation gives
\begin{equation}
\left(
\Phi_2'
-g_2\varphi^2
-\frac{\chi_2}{2}
\right)'
=
\frac{q^2(1-u)}{u^2f_0}\varphi^2.
\label{eq:sc_Phi2}
\end{equation}
For a given critical zero mode, Eqs.~\eqref{eq:sc_chi2} and
\eqref{eq:sc_Phi2} uniquely determine $\chi_2$ and $\Phi_2$.

The remaining Einstein equation determines the metric correction,
\begin{equation}
\begin{split}
f_2'
-\frac{3}{u}f_2
={}&
\frac12f_0\chi_2'
-\frac{\varphi^2}{2u}
+
\frac{u^3}{4}
\left[
2z_cz_2
+
z_c^2
\left(
\chi_2
+g_2\varphi^2
-2\Phi_2'
\right)
\right].
\end{split}
\label{eq:sc_f2}
\end{equation}
At this order, the horizon shift $z_2$ remains undetermined. We
therefore write
\begin{equation}
f_2
=
f_2^{(0)}
+
z_2 f_2^{(h)},
\qquad
f_2^{(h)}
=
\frac{z_c}{2}u^3(u-1),
\label{eq:sc_f2_split}
\end{equation}
where $f_2^{(0)}$ is determined by the critical zero mode, while
$f_2^{(h)}$ describes the response to a shift of the horizon radius.

At third order, the
scalar equation takes the inhomogeneous form
\begin{equation}
{\cal L}_c\phi_3
=
{\cal S}_3,
\label{eq:sc_cubic_equation}
\end{equation}
where
\begin{align}
{\cal S}_3
={}&
-\left[
\frac{f_2-\chi_2f_0/2}{u^2}\varphi'
\right]'
+
\left[
\frac{\chi_2}{u^4}
-g_2z_c^2
\left(
\frac{\chi_2}{2}
+\frac{2z_2}{z_c}
-2\Phi_2'
\right)
\right]\varphi
+
\left(
\frac{4v_4}{u^4}
-2g_4z_c^2
\right)\varphi^3
\nonumber\\
&-
\frac{z_c^2\Phi_0}{u^2f_0}
\left\{
q^2
\left[
2\Phi_2
+
\left(
\frac{\chi_2}{2}
-\frac{f_2}{f_0}
+\frac{2z_2}{z_c}
\right)\Phi_0
\right]\varphi
+
2\eta_4\Phi_0\varphi^3
\right\}.
\label{eq:sc_cubic_source}
\end{align}
Because ${\cal L}_c$ has the zero mode $\varphi$, the inhomogeneous
equation for $\phi_3$ admits a regular solution only if its source is
orthogonal to $\varphi$. This gives the Fredholm solvability condition
\begin{equation}
\int_0^1 du\,\varphi(u)\,{\cal S}_3(u)=0.
\label{eq:sc_fredholm}
\end{equation}
Since ${\cal S}_3$ depends on the undetermined horizon shift $z_2$,
Eq.~\eqref{eq:sc_fredholm} fixes $z_2$.

The temperature along the superconducting solution is expanded as
\begin{equation}
T
=
T_c
+
T_2\epsilon^2
+
T_4\epsilon^4
+
\mathcal O(\epsilon^6),
\label{eq:sc_T_expansion}
\end{equation}
where
\begin{equation}
T_2
=
-\frac{1}{4\pi z_c}
\left[
f_2'(1)
-\frac12f_0'(1)\chi_2(1)
-\frac{z_2}{z_c}f_0'(1)
\right].
\label{eq:sc_T2}
\end{equation}
For an ordinary critical point, $T_2\neq0$. On a superconducting branch
that emerges below $T_c$, one has $T_2<0$, and hence
\begin{equation}
T_c-T
=
-T_2\epsilon^2
+\mathcal O(\epsilon^4),
\qquad
O
=
\frac{1}{z_c^2\sqrt{-T_2}}
(T_c-T)^{1/2}
+\cdots.
\end{equation}
Thus an ordinary critical point exhibits the mean-field
scaling $O\propto(T_c-T)^{1/2}$.

\subsubsection{Representative models}
\label{subsec:ordinary_models}

We now specialize to three nonlinear completions,
\begin{equation}
\begin{split}
{\cal M}_{\rm AdS}:&\qquad
G(\phi)=1,
\qquad
V(\phi)=-6-\phi^2+\frac13\phi^6,
\qquad
J(\phi)=9\phi^2,
\\
{\cal M}_{\rm Lif}:&\qquad
G(\phi)=1,
\qquad
V(\phi)=-6-\phi^2+\frac{13475}{486}\phi^6,
\qquad
J(\phi)=9\phi^2,
\\
{\cal M}_{\rm run}:&\qquad
G(\phi)=1,
\qquad
V(\phi)=-6-\phi^2,
\qquad
J(\phi)=9\phi^2.
\end{split}
\label{eq:three_IR_completions}
\end{equation}
These models have identical couplings through quartic order around
$\phi=0$ and therefore share the same critical zero mode and cubic
solvability condition. They differ only in their higher-order
nonlinear completion, leading to the distinct zero-temperature
infrared endpoints discussed in Appendix~\ref{app:zero_temperature}.

The lowest nodeless zero mode determines the common critical data,
\begin{equation}
z_c=1.3066,
\qquad
T_c=0.15672.
\label{eq:critical_data}
\end{equation}
Solving the quadratic backreaction equations and imposing the Fredholm
condition gives
\begin{equation}
z_2=0.18454,
\qquad
T_2=-0.019137.
\end{equation}
The three models therefore share the same leading near-critical behavior,
\begin{equation}
T_c-T
=
0.019137\,\epsilon^2
+\mathcal O(\epsilon^4),
\qquad
O
\simeq
4.2343\,(T_c-T)^{1/2}.
\label{eq:sc_ordinary_scaling}
\end{equation}

\subsection{Multicritical extension}
\label{subsec:sc_tricritical}

To construct a tricritical point, we tune the nonlinear couplings so
that the quadratic temperature coefficient $T_2$ vanishes, while keeping
the critical RN background and scalar zero mode unchanged. Starting from
${\cal M}_{\rm AdS}$, we introduce the quartic St\"uckelberg coupling
\begin{equation}
G(\phi)=1,
\qquad
V(\phi)=-6-\phi^2+\frac13\phi^6,
\qquad
J(\phi)=9\phi^2+\eta_4\phi^4.
\end{equation}
Since $\eta_4$ first enters the cubic source
\eqref{eq:sc_cubic_source}, the critical zero mode and the quadratic
backreaction are unchanged. The Fredholm condition gives
\begin{equation}
T_2(\eta_4)
=
-0.019137
+
0.00071792\,\eta_4.
\end{equation}
The tricritical point is therefore located at
$\eta_4^t=26.656$, where $T_2(\eta_4^t)=0$.

At this tuned point, the fifth-order solvability condition gives
$T_4=-0.0053082$, leading to
\begin{equation}
T_c-T
=
0.0053082\,\epsilon^4
+\mathcal O(\epsilon^6),
\qquad
O
\simeq
2.1701\,(T_c-T)^{1/4}.
\label{eq:tricritical_scaling}
\end{equation}

More generally, if independent nonlinear couplings are tuned such that
\begin{equation}
T_2=T_4=\cdots=T_{2r-2}=0,
\qquad
T_{2r}<0,
\end{equation}
then
\begin{equation}
T_c-T
\propto
\epsilon^{2r},
\qquad
O
\propto
(T_c-T)^{1/(2r)}.
\label{eq:multicritical_amplitude_law}
\end{equation}
Thus multicritical tuning changes the amplitude law while leaving the
critical RN background and linear zero mode unchanged.

\subsection{Near-critical thermodynamics}
\label{subsec:near_critical_thermo}

We now consider the near-critical thermodynamics. At fixed chemical
potential, we compare the hairy solution with the RN solution at the
same temperature through
\begin{equation}
\Delta{\cal F}
\equiv
{\cal F}_{\rm RN}(T)
-
{\cal F}_{\rm hairy}(T),
\label{eq:deltaF_definition}
\end{equation}
A positive $\Delta{\cal F}$ therefore indicates that the hairy solution
is thermodynamically preferred.

Using the first law at fixed $\mu$,
$d{\cal F}=-s\,dT$, we have
\begin{equation}
\frac{d\Delta{\cal F}}{d\epsilon}
=
\left(
s_{\rm hairy}-s_{\rm RN}
\right)
\frac{dT}{d\epsilon}.
\label{eq:deltaF_first_law}
\end{equation}
For the ordinary critical branch, matching the RN solution to the same
temperature gives
\begin{equation}
z_h^{\rm RN}
=
z_c
+
\widehat z_2\epsilon^2
+
\mathcal O(\epsilon^4),
\qquad
\widehat z_2
=
\frac{T_2}{T'_{\rm RN}(z_c)}
=
0.11981.
\label{eq:RN_horizon_shift}
\end{equation}
Together with $z_2=0.18454$ for the hairy branch and
$s=4\pi/z_h^2$, this gives
\begin{equation}
s_{\rm hairy}-s_{\rm RN}
=
-0.72934\,\epsilon^2
+
\mathcal O(\epsilon^4).
\label{eq:entropy_difference}
\end{equation}
Combining this result with  \eqref{eq:sc_ordinary_scaling} gives
\begin{equation}
\Delta{\cal F}
\simeq
19.056\,(T_c-T)^2.
\label{eq:deltaF_temperature}
\end{equation}

For the tricritical point, the cubic solvability condition gives
$z_2=0.064734$, and hence
\begin{equation}
s_{\rm hairy}-s_{\rm RN}
=
-0.72934\,\epsilon^2
+\mathcal O(\epsilon^4).
\end{equation}
Using \eqref{eq:tricritical_scaling} in
\eqref{eq:deltaF_first_law}, we obtain
\begin{equation}
\Delta{\cal F}
\simeq
6.6737\,(T_c-T)^{3/2}.\label{eq:deltaF_tri}
\end{equation}

\begin{figure}[h!]
\centering
\includegraphics[width=0.8\textwidth]{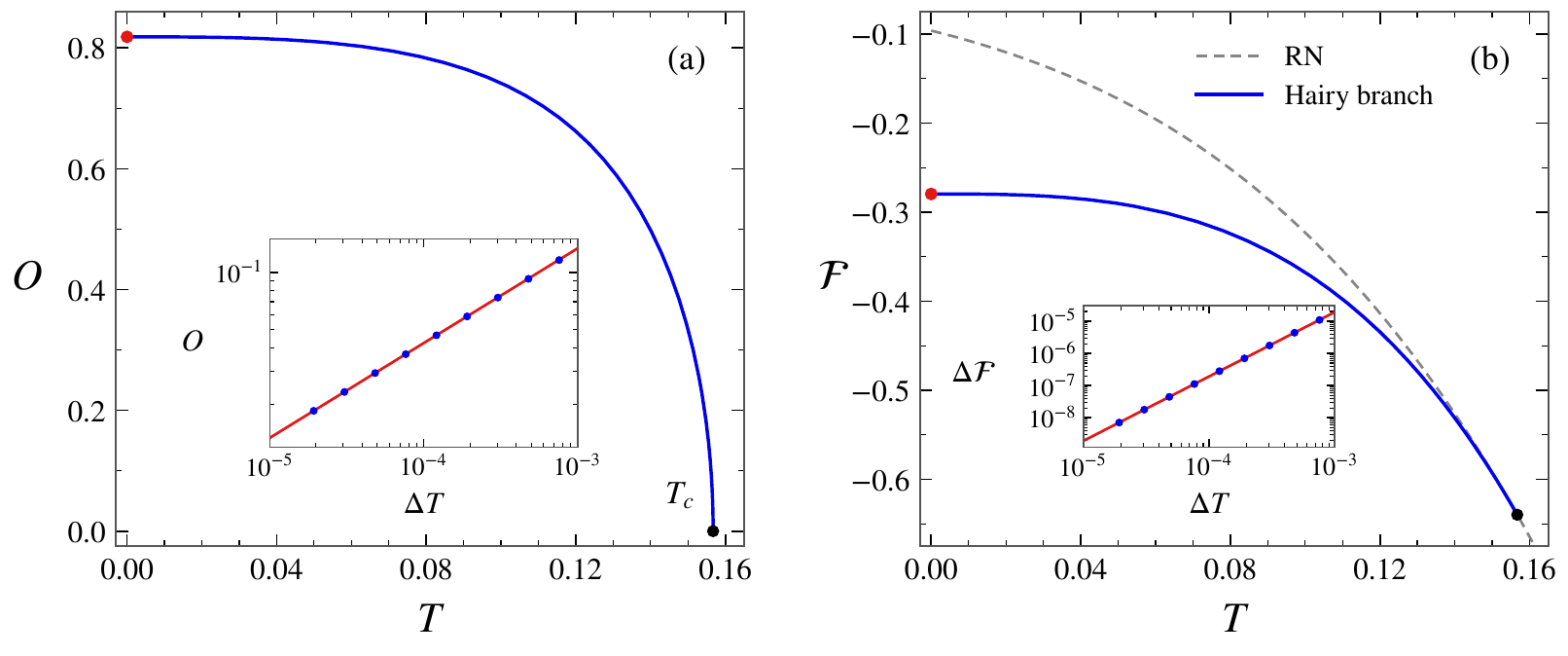}
\caption{\small
Condensate and grand potential for $\mathcal M_{\rm AdS}$.
Panel (a) shows the condensate as a function of temperature, while
panel (b) compares the hairy and RN branches.
The black points mark the critical temperature $T_c$, and the red points
denote the zero-temperature solutions. The insets compare the numerical solutions with the near-critical
predictions in Eqs.~\eqref{eq:sc_ordinary_scaling} and
\eqref{eq:deltaF_temperature}; the red lines denote the analytic
asymptotics.
}
\label{fig:num_MF_thermo}
\end{figure}

\begin{figure}[h!]
\centering
\includegraphics[width=0.8\textwidth]{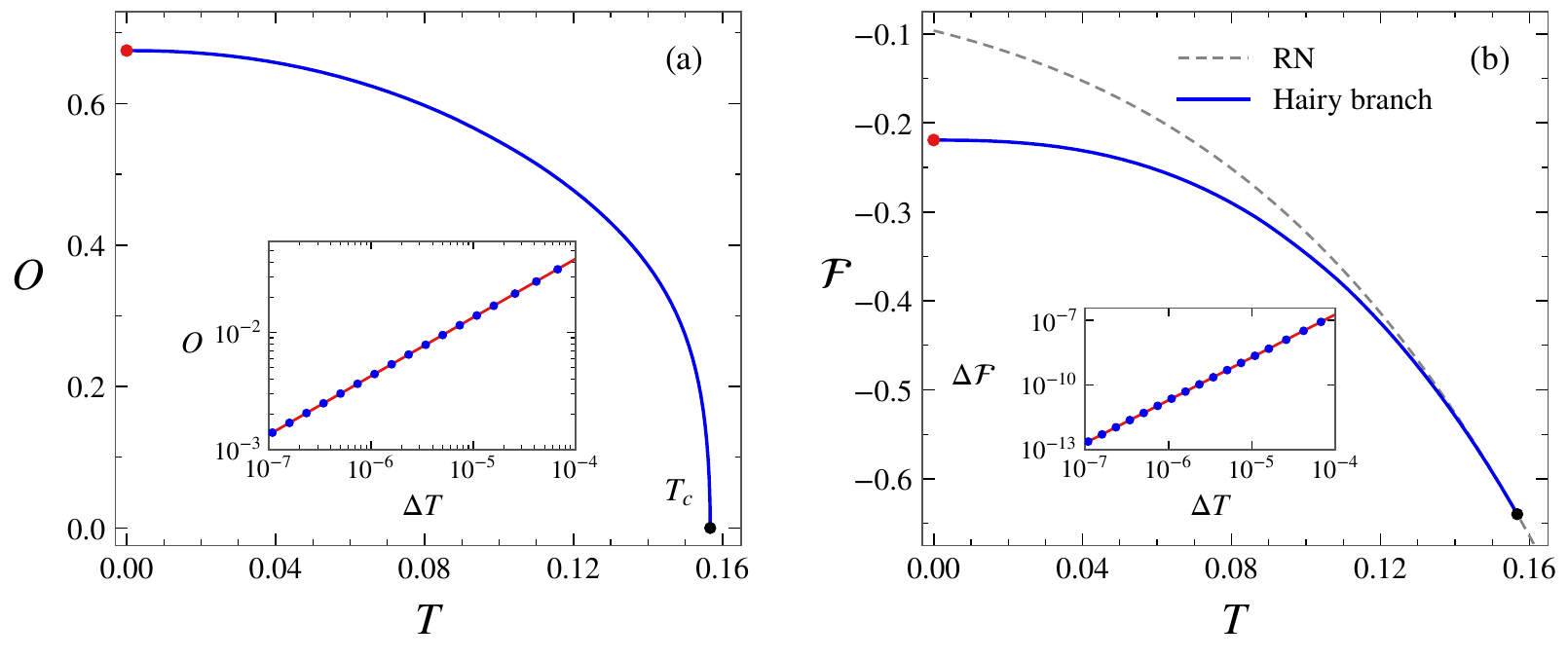}
\caption{\small
Condensate and grand potential for $\mathcal M_{\rm Lif}$.
Panel (a) shows the condensate as a function of temperature, while
panel (b) compares the hairy and RN branches.
The black points mark the critical temperature $T_c$, and the red points
denote the zero-temperature solutions. The insets compare the numerical solutions with the near-critical
predictions in Eqs.~\eqref{eq:sc_ordinary_scaling} and
\eqref{eq:deltaF_temperature}; the red lines denote the analytic
asymptotics.
}
\label{fig:num_Lif_thermo}
\end{figure}

\begin{figure}[h!]
\centering
\includegraphics[width=0.8\textwidth]{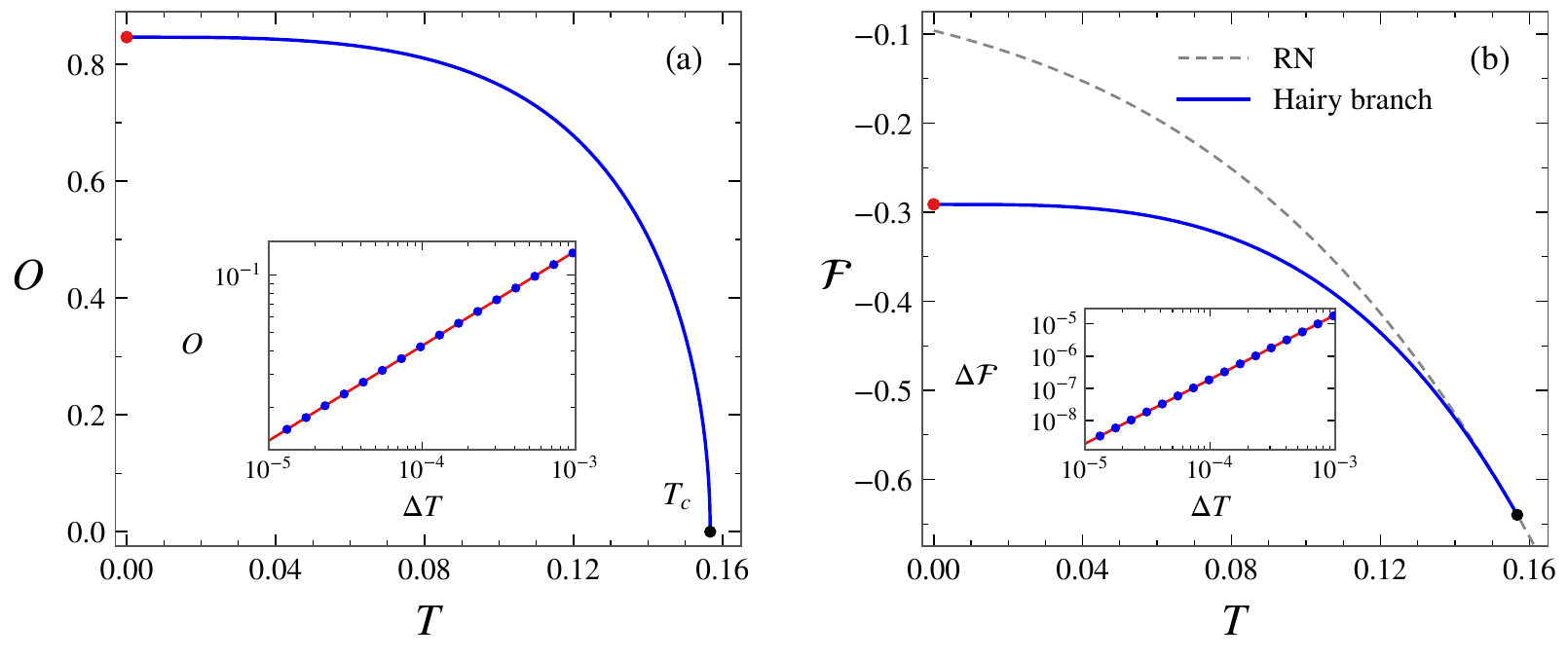}
\caption{\small
Condensate and grand potential for $\mathcal M_{\rm run}$.
Panel (a) shows the condensate as a function of temperature, while
panel (b) compares the hairy and RN branches.
The black points mark the critical temperature $T_c$, and the red points
denote the zero-temperature solutions. The insets compare the numerical solutions with the near-critical
predictions in Eqs.~\eqref{eq:sc_ordinary_scaling} and
\eqref{eq:deltaF_temperature}; the red lines denote the analytic
asymptotics.
}
\label{fig:num_Run_thermo}
\end{figure}

\begin{figure}[h!]
\centering
\includegraphics[width=0.8\textwidth]{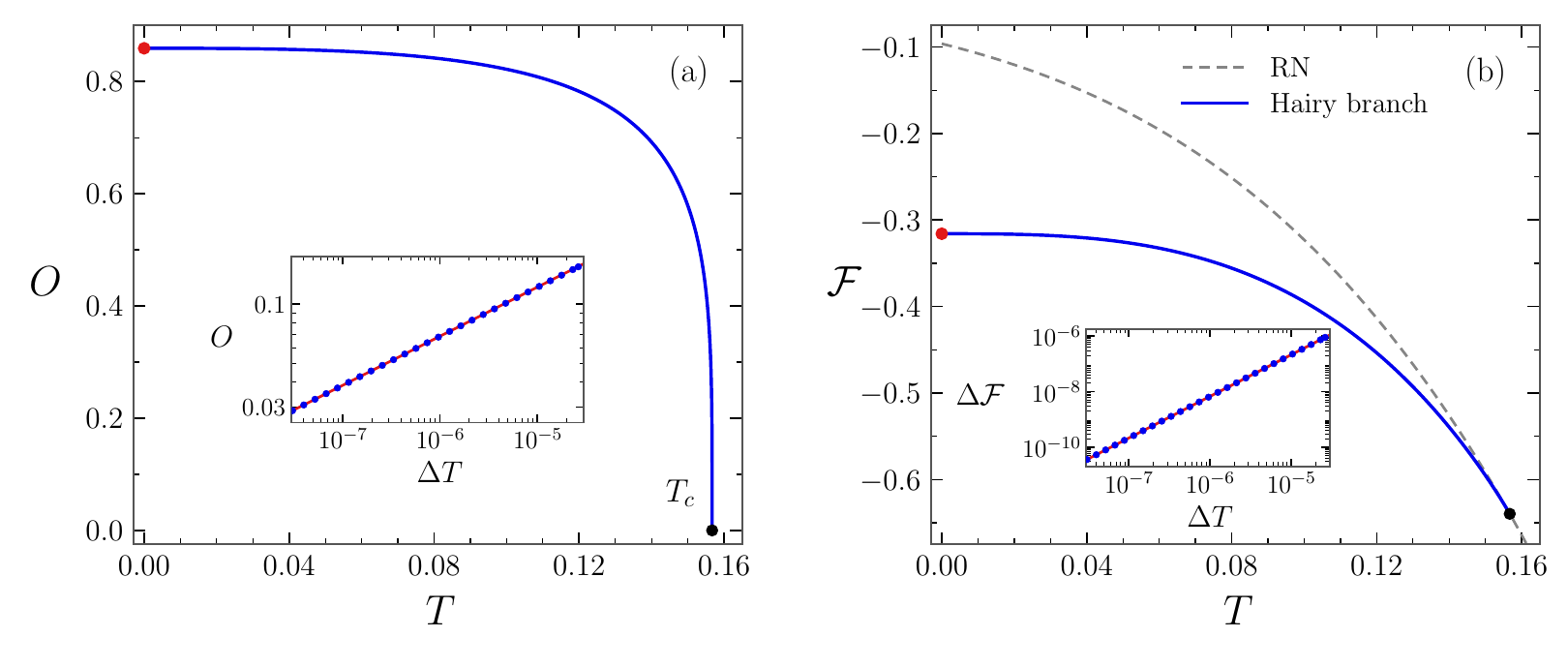}
\caption{\small
Condensate and grand potential for $\mathcal M_{\rm AdS}$ at the
tricritical coupling $\eta_4=\eta_4^t$.
Panel (a) shows the condensate as a function of temperature, while
panel (b) compares the hairy and RN branches.
The black points mark the critical temperature $T_c$, and the red points
denote the zero-temperature solutions. The insets compare the numerical solutions with the tricritical
predictions in Eqs.~\eqref{eq:tricritical_scaling} and
\eqref{eq:deltaF_tri}; the red lines denote the analytic asymptotics.
}
\label{fig:num_tri_thermo}
\end{figure}

The thermodynamic behavior of the representative cases is summarized in
Figs.~\ref{fig:num_MF_thermo}--\ref{fig:num_tri_thermo}.
For the three ordinary critical models, the condensate and grand potential
confirm the common near-critical scaling derived above, while their behavior
away from $T_c$ depends on the nonlinear completion. The tricritical case
instead exhibits the modified scaling associated with the tuned quartic
St\"uckelberg coupling.

The thermodynamic analysis thus fixes the exterior critical behavior
whose transmission into the black-hole interior is studied next.

\section{Transmission from the critical zero mode to the black-hole interior}
\label{sec:interior}

We now follow the critical scalar zero mode into the black-hole interior.
At the critical point, the reference RN--AdS$_4$ solution possesses a
regular Cauchy horizon. Turning on an arbitrarily small condensate,
however, produces a nonuniform perturbation in its vicinity. The
critical zero mode becomes logarithmically oscillatory near the
reference Cauchy horizon, where its enhanced radial gradient drives the
geometry into an Einstein--Rosen collapse. The solution subsequently
enters the Josephson regime and eventually approaches the Kasner
epochs.

We first analyze this transmission at fixed small condensate amplitude
$\epsilon$. Since the ordinary and tricritical solutions share the same critical
RN geometry and linear zero mode, their leading interior transmission
is identical at fixed $\epsilon$. The distinction appears only after
converting from $\epsilon$ to the reduced temperature
$t\equiv1-T/T_c$.

\subsection{From the critical zero mode to Einstein--Rosen collapse}
\label{subsec:CH_transmission}

The critical zero mode obtained in
Sec.~\ref{subsec:critical_bifurcation} can be continued
through the event horizon into the RN interior. The reference geometry
has a regular Cauchy horizon at $z=z_-$. In its vicinity
\begin{equation}
f_c(z)
=
f'_-(z-z_-)
+
O\!\left((z-z_-)^2\right),
\qquad
\Phi_c(z)
=
\Phi_-+O(z-z_-).
\label{eq:CH_background}
\end{equation}
where $f'_->0$, and $\Phi_-$ is the electrostatic potential at the
Cauchy horizon, with the gauge fixed by $\Phi(z_c)=0$ at the event
horizon.

Near the Cauchy horizon, the charged-scalar term acquires a double-pole
enhancement, $f_c^{-2}\sim(z-z_-)^{-2}$, while the mass and other
coupling terms remain regular. The critical zero mode therefore obeys,
at leading order,
\begin{equation}
\varphi''
+
\frac{1}{z-z_-}\varphi'
+
\frac{\nu_-^2}{(z-z_-)^2}\varphi
\simeq0,
\qquad
\nu_-
=
\frac{|q\Phi_-|}{f'_-}.
\label{eq:CH_scalar_equation}
\end{equation}
A Frobenius ansatz $\phi\sim(z-z_-)^\lambda$ gives
$\lambda=\pm i\nu_-$, implying oscillatory behavior in
$\log|z-z_-|$. The mode that is regular at the event horizon therefore
takes the near-Cauchy-horizon form
\begin{equation}
\varphi\left(\frac{z}{z_c}\right)
=
{\cal T}_-\cos\Theta_-+\cdots,
\qquad
\Theta_-
=
\nu_-
\log\left|
\frac{z-z_-}{\ell_-}
\right|
+
\delta_- .
\label{eq:CH_solution}
\end{equation}
The amplitude ${\cal T}_-$ and phase $\delta_-$ are fixed by continuing
the normalized critical zero mode from the event horizon to the Cauchy
horizon, while $\ell_-$ only sets the phase convention. For the critical
RN solution, the transmission amplitude is
\begin{equation}
\nu_-=0.63304,
\qquad
{\cal T}_-=0.50227.
\label{eq:CH_benchmark}
\end{equation}

On the superconducting solution, the transmitted scalar has amplitude
${\cal A}_-=\epsilon{\cal T}_-$ and therefore behaves as
\begin{equation}
\phi
=
{\cal A}_-\cos\Theta_-+\cdots .
\label{eq:CH_physical}
\end{equation}
Although the scalar remains $O(\epsilon)$, its radial derivative is enhanced near the
Cauchy horizon as $\phi'\sim\epsilon/(z-z_-)$. The resulting leading
contribution to the redshift function is
\begin{equation}
\chi'
\simeq
\frac{z\nu_-^2{\cal A}_-^2}
{2(z-z_-)^2}.
\label{eq:redshift_enhancement}
\end{equation}
Integrating this behavior gives
$\delta\chi\sim\epsilon^2/|z-z_-|$. The perturbative expansion around
the RN interior therefore ceases to be valid when
$|z-z_-|\sim\epsilon^2$. At this scale the scalar amplitude remains
parametrically small, while its backreaction on the geometry becomes
order unity and triggers the nonlinear Einstein--Rosen collapse.

To describe the collapse layer, we denote its characteristic radial
location by the would-be inner horizon $z_\star$. We further introduce
the local background quantities
\begin{equation}
\Phi_o \equiv \Phi(z_\star),
\qquad
D_o \equiv z_\star^4 E_o^2-12,
\end{equation}
where $E_o$ is the local electric field evaluated at $z_\star$. In the critical limit, the collapse layer is attached to the RN inner horizon, so that
$z_\star\to z_-$, $\Phi_o\to\Phi_-$, and $D_o\to4z_-f'_-$.

The local scalar solution across the outer side of the collapse layer,
derived in Appendix~\ref{subsec:ER_matching}, takes the form
\begin{equation}
\phi
=
\phi_o
\cos\left(
q\Phi_o\int^z\frac{e^{\chi/2}}{f}\,dz'
+\varphi_o
\right).
\end{equation}
Using the RN overlap introduced above, its phase reduces to
\begin{equation}
q\Phi_o\int^z\frac{e^{\chi/2}}{f}\,dz'\simeq
\frac{q\Phi_-}{f'_-}\log|z-z_-|
=
\pm\nu_-\log|z-z_-|,
\label{eq:ER_phase_matching}
\end{equation}
which reproduces the logarithmic phase of the transmitted critical mode.
Matching the amplitude then gives
\begin{equation}
\phi_o
=
{\cal A}_-
\label{eq:ER_scalar_matching}
\end{equation}

With the scalar amplitude fixed, we now match the metric sector of the
local ER solution derived in Appendix~\ref{subsec:ER_matching}. The local
constants $c_1$ and $c_2$ are defined in \eqref{eq:ER_c1c2}, while
the characteristic width of the exponential collapse is
$\Delta z_{\rm ER}=c_1^2/c_2^2$ from
\eqref{eq:ER_width_general}.

The outer side of the ER layer must reproduce the linear RN geometry.
Using ${\cal G}\simeq-f'_-(z-z_-)/z_-^2$ for the RN solution and
${\cal G}'\simeq-c_2^2$ from
\eqref{app:eq:ER_first_order}, matching the slopes gives
$c_2^2=f'_-/z_-^2\,[1+o(1)]$. Together with
\eqref{eq:ER_c1c2}, \eqref{eq:ER_scalar_matching} and the critical limits introduced above, this yields
\begin{equation}
\Delta z_{\rm ER}
=
\frac{z_-\nu_-^2}{4}\,
{\cal A}_-^2
=
0.10820\,\epsilon^2
\label{eq:ER_width}
\end{equation}

\subsection{Josephson matching and Kasner transfer}
\label{subsec:Josephson}

The ER collapse fixes the radial scale of the subsequent Josephson regime.
Using the local matching relation
\eqref{app:eq:c3_matching} together with the RN--ER matching obtained
in the previous subsection, we find that near the critical point
\begin{equation}
c_3
=
\frac{z_-D_-}{q^2\Phi_-^2{\cal A}_-^2},
\label{eq:Josephson_c3}
\end{equation}
where $D_-=4z_-f'_-$. Thus the Josephson radial scale diverges as
$c_3\propto{\cal A}_-^{-2}$ near criticality.

The ER solution matches onto the large-$x$ oscillatory branch of
the Bessel solution \eqref{app:eq:Josephson_solution}. Using the local matching point
\eqref{app:eq:Josephson_xstar_local}, its critical behavior is
\begin{equation}
x_\star
=
\frac{2}{\nu_-{\cal A}_-^2}.
\label{eq:Josephson_xstar}
\end{equation}

Using \eqref{app:eq:Josephson_amplitude_local}, the Josephson amplitude becomes
\begin{equation}
A_-
=
\sqrt{\frac{\pi}{\nu_-}}.
\label{eq:Josephson_amplitude}
\end{equation}
Thus the oscillation amplitude remains finite in the critical limit and
is fixed entirely by the inner-horizon data.

At the matching point, the large-$x$ Josephson solution carries the
phase $x_\star-\pi/4-\Theta_J$. Since the phase inherited from the ER
side remains finite, the divergent part of $\Theta_J$ must coincide
with that of $x_\star$. Using \eqref{eq:Josephson_xstar} and
${\cal A}_-=\epsilon{\cal T}_-$, we therefore obtain
\begin{equation}
\Theta_J
=\frac{2}{\nu_-{\cal A}_-^2}+
C_-^J
+
o(1)=
\frac{B_-^J}{\epsilon^2}
+
C_-^J
+
o(1),
\qquad
B_-^J
=
\frac{2}{\nu_-{\cal T}_-^2}.
\label{eq:Josephson_phase}
\end{equation}
Here $C_-^J$ denotes the finite phase offset left after subtracting the
universal divergent contribution. For the critical inner horizon data in
\eqref{eq:CH_benchmark},
\begin{equation}
A_-=2.2277,
\qquad
B_-^J=12.523.
\label{eq:Josephson_benchmark}
\end{equation}

The quartic St\"uckelberg interaction produces a logarithmic correction
to the Josephson phase. Using the local phase shift derived in
\eqref{app:eq:Josephson_frequency_shift}, we obtain
\begin{equation}
\Theta_J
=
\frac{B_-^J}{\epsilon^2}
+
\kappa\log\left(\epsilon^{-2}\right)
+
C_-^J
+
o(1),
\label{eq:Josephson_phase_general}
\end{equation}
where
\begin{equation}
\kappa
=
\frac{3\eta_4}{2q^2\nu_-}.
\label{eq:Josephson_kappa}
\end{equation}
Thus the leading $\epsilon^{-2}$ phase accumulation is fixed by the
linear transmission data, while the quartic interaction enters first
through the logarithmic correction.

As the evolution proceeds inward, the Josephson solution enters its
small-$x$ regime. Using \eqref{app:eq:Josephson_velocity}, we obtain
\begin{equation}
v_1
\equiv
\lim z\phi'
=
-\frac{4c_5}{\pi}
=
-\frac{4A_-}{\pi}\sin\Theta_J
+
o(1).
\label{eq:Kasner_velocity_phase}
\end{equation}
The Josephson phase is thus transferred directly into the scalar
velocity that initializes the first Kasner epoch.

For later convenience, we define
$\alpha_0
\equiv
v1/2$.
In the quadratic approximation, \eqref{eq:Kasner_velocity_phase} then gives
\begin{equation}
\alpha_0
=
\alpha_A\sin\Theta_J
+
o(1),
\qquad
|\alpha_A|
=
\frac{2}{\sqrt{\pi\nu_-}}
=
1.4182.
\label{eq:alpha_transfer}
\end{equation}

Once $z\phi'$ approaches the constant $v_1$, the leading interior
equations give
\begin{equation}
\phi
\simeq
v_1\log z+\mathrm{const.},
\qquad
\chi
\simeq
\frac{v_1^2}{2}\log z+\mathrm{const.},
\qquad
f
\simeq
-f_1 z^{3+v_1^2/4},
\end{equation}
where $f_1>0$ is a constant. Transforming to the proper-time Kasner form,
the corresponding Kasner exponents are
\begin{equation}
p_t^{(1)}
=
\frac{v_1^2-4}{v_1^2+12},
\qquad
p_x^{(1)}
=
p_y^{(1)}
=
\frac{8}{v_1^2+12},
\qquad
p_\phi^{(1)}
=
\frac{4\sqrt{2}\,v_1}{v_1^2+12}.
\label{eq:first_Kasner_exponents}
\end{equation}

More general nonlinear interactions can change the functional dependence
of $v_1$ on $\Theta_J$. We therefore parameterize the leading phase transfer by
\begin{equation}
v_1
=
{\cal V}(\Theta_J)
+
o(1),
\label{eq:Kasner_transfer_function}
\end{equation}
where ${\cal V}$ is the nonlinear transfer function from the
Josephson phase to the first-Kasner scalar velocity. The
$\phi\to-\phi$ symmetry implies
\begin{equation}
{\cal V}(\Theta+\pi)
=
-{\cal V}(\Theta).
\label{eq:transfer_symmetry}
\end{equation}
This implies that ${\cal V}$ changes
sign under a phase shift of $\pi$. Hence, for continuous ${\cal V}$,
there is at least one zero in every interval of length $\pi$.
For the accumulation analysis below, we assume that the relevant zero
is simple,
\begin{equation}
{\cal V}(\Theta_\star)=0,
\qquad
{\cal V}'(\Theta_\star)\neq0.
\label{eq:transfer_simple_zero}
\end{equation}
Under this assumption, the subleading corrections only shift a simple zero perturbatively.
The accumulation condition is therefore determined by
$\Theta_J\to\Theta_\star$.

Equivalently, the transfer relation can be written as
\begin{equation}
\alpha_0
=
{\cal F}(\Theta_J)
+
o(1),
\qquad
{\cal F}(\Theta)
\equiv
\frac{1}{2}{\cal V}(\Theta).
\label{eq:Kasner_transfer_alpha}
\end{equation}
At the tricritical point, $\eta_4=\eta_4^t$, the quartic interaction
modifies the quadratic sinusoidal transfer, so that ${\cal F}(\Theta)$ must
in general be determined numerically.

\subsection{Ordinary and tricritical scaling}
\label{subsec:Kasner_scaling}

We now translate the near-critical interior scaling from the amplitude
$\epsilon$ to the reduced temperature
$t\equiv1-T/T_c$. Suppose that the first nonvanishing correction to the
temperature occurs at order $\epsilon^{2r}$,
\begin{equation}
t
=
a_{2r}\epsilon^{2r}
+
o(\epsilon^{2r}),
\qquad
a_{2r}
=
-\frac{T_{2r}}{T_c}
>
0.
\label{eq:temperature_scaling_general}
\end{equation}
Using \eqref{eq:ER_width}, the ER-collapse width becomes
\begin{equation}
\Delta z_{\rm ER}
=
\frac{z_-\nu_-^2{\cal T}_-^2}{4}\,
a_{2r}^{-1/r}t^{1/r}.
\label{eq:temperature_ER_width}
\end{equation}
Likewise, Eq.~\eqref{eq:Josephson_phase_general} gives
\begin{equation}
\Theta_J(t)
=
B_r t^{-1/r}
+
\kappa\log\left(t^{-1/r}\right)
+
C_r
+
o(1),
\qquad
B_r
=
\frac{2a_{2r}^{1/r}}{\nu_-{\cal T}_-^2}.
\label{eq:temperature_phase_relation}
\end{equation}
The finite contribution generated by $\log a_{2r}^{1/r}$ has been absorbed into $C_r$.

Since the first-Kasner exponents are determined by the Josephson phase,
they inherit the same rapidly varying dependence on $t$,
\begin{equation}
p_i^{(1)}(t)
=
P_i\!\left(\Theta_J(t)\right)
+
o(1).
\label{eq:Kasner_periodic_scaling}
\end{equation}
Let $\Theta_\star$ denote a simple zero of the transfer function.
Successive zeros then satisfy
\begin{equation}
B_r t_n^{-1/r}
+
\kappa\log\left(t_n^{-1/r}\right)
+
C_r
=
\Theta_\star+n\pi
+
o(1).
\label{eq:Kasner_zero_phase_condition}
\end{equation}
At large $n$,
\begin{equation}
t_n
=
\left(
\frac{B_r}{\pi n}
\right)^r
\left[
1+
O\left(
\frac{\log n}{n}
\right)
\right],
\label{eq:sec4_root_asymptotic}
\end{equation}
and hence
\begin{equation}
t_n\sim n^{-r}.
\label{eq:Kasner_zero_accumulation}
\end{equation}
Thus the order at which the temperature first departs from $T_c$
directly determines the accumulation exponent of the interior
oscillations.

\subsubsection{Ordinary critical point}

For an ordinary critical point,
Eq.~\eqref{eq:sc_ordinary_scaling} gives
\begin{equation}
r=1,
\qquad
a_2=0.12211.
\end{equation}
Since $\eta_4=0$, the logarithmic correction is absent, $\kappa=0$.
Eq.~\eqref{eq:temperature_ER_width} and
\eqref{eq:temperature_phase_relation} therefore reduce to
\begin{equation}
\Delta z_{\rm ER}
\simeq
0.88606\,t,
\qquad
\Theta_J(t)
=
\frac{1.5292}{t}
+
C_1
+
o(1).
\label{eq:temperature_ER_widthnum}
\end{equation}

For the quadratic transfer,
Eq.~\eqref{eq:alpha_transfer} gives the explicit first-Kasner waveform
\begin{equation}
\begin{split}
p_t^{(1)}(t)
&=
\frac{
\alpha_A^2\sin^2\Theta_J(t)-1
}{
\alpha_A^2\sin^2\Theta_J(t)+3
},
\\
p_x^{(1)}(t)
=
p_y^{(1)}(t)
&=
\frac{
2
}{
\alpha_A^2\sin^2\Theta_J(t)+3
},
\\
p_\phi^{(1)}(t)
&=
\frac{
2\sqrt{2}\,\alpha_A\sin\Theta_J(t)
}{
\alpha_A^2\sin^2\Theta_J(t)+3
}.
\end{split}
\label{eq:sec4_explicit_waveform}
\end{equation}
The metric exponents are $\pi$-periodic in $\Theta_J$, while the scalar
velocity changes sign under $\Theta_J\to\Theta_J+\pi$. The corresponding
zeros accumulate as
\begin{equation}
t_n\sim n^{-1}.\label{eq:Kasner_zero_accumulation2nd}
\end{equation}

\subsubsection{Tricritical point}

At the tricritical point,
Eq.~\eqref{eq:tricritical_scaling} gives
\begin{equation}
r=2,
\qquad
a_4
=
0.033871.
\end{equation}
The ER-collapse width and Josephson phase become
\begin{equation}
\Delta z_{\rm ER}
\simeq
0.58790\,t^{1/2},
\qquad
\Theta_J(t)
=
2.3048\,t^{-1/2}
+
7.0181\,\log\left(t^{-1/2}\right)
+
C_2
+
o(1).\label{eq:temperature_ER_widthnumtri}
\end{equation}

Using \eqref{eq:Kasner_transfer_alpha}, the first-Kasner exponents are
\begin{equation}
\begin{split}
p_t^{(1)}(t)
&=
\frac{
{\cal F}^2\!\left(\Theta_J(t)\right)-1
}{
{\cal F}^2\!\left(\Theta_J(t)\right)+3
},
\\
p_x^{(1)}(t)
=
p_y^{(1)}(t)
&=
\frac{
2
}{
{\cal F}^2\!\left(\Theta_J(t)\right)+3
},
\\
p_\phi^{(1)}(t)
&=
\frac{
2\sqrt{2}\,{\cal F}\!\left(\Theta_J(t)\right)
}{
{\cal F}^2\!\left(\Theta_J(t)\right)+3
}.
\end{split}
\label{eq:tricritical_Kasner_waveform}
\end{equation}

Provided the relevant zero of ${\cal F}$ remains simple,
Eq.~\eqref{eq:Kasner_zero_accumulation} gives
\begin{equation}
t_n\sim n^{-2}.
\label{eq:tricritical_accumulation}
\end{equation}
The logarithmic phase correction affects only the subleading spacing of
the accumulation sequence.

\subsection{Electric Kasner inversion}
\label{subsec:Kasner_inversion}

The first Kasner epoch can be destabilized deeper in the interior when
the residual electric field becomes important. For the constant Maxwell
coupling considered here, the resulting transition is the electric
Kasner inversion of Ref.~\cite{Hartnoll:2020fhc}.

When the electric flux is approximately conserved across the transition
and the scalar potential and St\"uckelberg forces remain subleading, the
two kinetic epochs satisfy
\begin{equation}
v_1v_2=4,
\qquad
p_t^{(2)}
=
-\frac{p_t^{(1)}}{1+2p_t^{(1)}}.
\label{eq:electric_Kasner_inversion}
\end{equation}
For $0<|v_1|<2$, or equivalently
$-1/3<p_t^{(1)}<0$, the inversion gives $|v_2|>2$ and hence
$p_t^{(2)}>0$.

The near-critical scaling derived above therefore determines the first
Kasner epoch, while any subsequent inversion depends on the nonlinear
interior evolution beyond the Josephson-to-Kasner transfer.

\section{Low-temperature Kasner scaling}
\label{sec:low_temperature}

We now turn to the low-temperature limit $T\to0$ for the three
representative models introduced in Sec.~\ref{subsec:ordinary_models}.
Holographic superconductors can approach distinct zero-temperature
infrared geometries~\cite{Gubser:2009ground,Horowitz:2009zero}.
The infrared solutions of our models are derived in
Appendix~\ref{app:zero_temperature} and lead to different
low-temperature interior scaling. The tricritical case
considered above corresponds to a particular parameter choice within
this set and does not define an additional infrared class.

We label successive kinetic epochs by $i$ and denote their scalar
velocities by $v_i=z\phi'$. The corresponding Kasner exponent is
\begin{equation}
p_t^{(i)}
=
\frac{v_i^2-4}{v_i^2+12}.
\label{eq:lowT_Kasner_map}
\end{equation}

\subsection{Cohesive \texorpdfstring{$AdS_4$}{AdS4} endpoint}
\label{subsec:lowT_AdS4}

For $\mathcal M_{\rm AdS}$, the zero-temperature solution approaches the
cohesive AdS$_4$ endpoint
$\phi_*=1$, $f_*=10/9$, with the leading irrelevant scalar deformation
\begin{equation}
\phi_0
=
\phi_*
-
A_\phi z^{-\delta_\phi}
+\cdots,
\qquad
\delta_\phi=1.5741,
\qquad
A_\phi=1.0020,
\end{equation}
as derived in \eqref{app:eq:adsirphi} and
\eqref{app:eq:AdS4_global_data}.

At small temperature, the leading IR geometry is obtained by replacing
the AdS$_4$ endpoint with an AdS--Schwarzschild horizon,
\begin{equation}
f
=
f_*
\left[
1-\left(\frac{z}{z_h}\right)^3
\right],
\qquad
\chi
\simeq
\chi_{\rm IR}.
\end{equation}
The temperature is therefore
\begin{equation}
Tz_h
=
{\cal T}_{\rm A}
\equiv
\frac{3f_*}{4\pi}
e^{-\chi_{\rm IR}/2}
=
0.22193.
\label{eq:lowT_AdS4_temperature}
\end{equation}

The irrelevant zero-temperature scalar mode must now be continued
through this thermal geometry. Writing $x=z/z_h$ and
$\delta\phi=C_hU(x)$, the linearized scalar equation is
\begin{equation}
U''
+
\left(
\frac{b'}{b}
-\frac{2}{x}
\right)U'
-
\frac{V_{,\phi\phi,*}}{f_*x^2b}\,U
=
0,
\qquad
b(x)=1-x^3.
\label{eq:lowT_AdS4_thermal_scalar}
\end{equation}
where primes denote derivatives with respect to $x$. Normalizing the horizon-regular solution by $U(1)=1$, its continuation
to the two asymptotic regions defines the coefficients $a_-$ and $w_K$ by
\begin{equation}
U(x)
\simeq
a_- x^{-\delta_\phi},
\qquad
x\ll1,
\qquad\qquad
xU'(x)
\longrightarrow
w_K,
\qquad
x\gg1.
\label{eq:lowT_AdS4_transfer}
\end{equation}
Here $a_-$ measures the amplitude of the exterior irrelevant mode, while
$w_K$ determines the scalar velocity transmitted into the first interior
kinetic region. Numerically,
\begin{equation}
a_-=1.2980,
\qquad
w_K=-0.95206.
\end{equation}

Matching the exterior tail to the zero-temperature solution fixes
$C_h=-A_\phi z_h^{-\delta_\phi}/a_-$. The first interior scalar velocity
is therefore
\begin{equation}
v_1
=
-\frac{A_\phi w_K}{a_-}
z_h^{-\delta_\phi}
=
{\cal V}_{\rm A}T^{\delta_\phi},
\qquad
{\cal V}_{\rm A}=7.8588.
\label{eq:lowT_AdS4_vint}
\end{equation}
Hence $v_1\to0$ and the first Kasner epoch approaches the Schwarzschild
value $p_t^{(1)}=-1/3$.

The residual electric field eventually triggers the electric Kasner
inversion of Sec.~\ref{subsec:Kasner_inversion}. Since $v_1v_2=4$, the
second velocity diverges as $T^{-\delta_\phi}$ and
$p_t^{(2)}\to1$. Expanding Eq.~\eqref{eq:lowT_Kasner_map} gives
\begin{equation}
p_t^{(1)}+\frac13
\simeq
6.8623\,T^{3.1482},
\qquad
1-p_t^{(2)}
\simeq
61.760\,T^{3.1482}.
\label{eq:lowT_AdS4_main_result}
\end{equation}
Both deviations are therefore governed by the same infrared exponent
$2\delta_\phi$.

\subsection{Finite-scalar Lifshitz endpoint}
\label{subsec:lowT_Lifshitz}

For $\mathcal M_{\rm Lif}$, the zero-temperature solution approaches the
charged Lifshitz fixed point
\begin{equation}
\phi_L=\frac{3}{\sqrt{35}},
\qquad
f_L=\frac{81}{140},
\qquad
\mathfrak z=2,
\end{equation}
derived in Eq.~\eqref{app:eq:Lifshitz_algebraic}. Unlike the cohesive
AdS$_4$ endpoint, the electric field remains part of the leading IR
geometry. The low-temperature limit must therefore be obtained from a
finite-temperature Lifshitz solution rather than from a neutral
blackening deformation.

To obtain the low-temperature thermal geometry, we introduce
$x=z/z_h$ and take the limit $z_h\to\infty$ at fixed $x$. In this limit,
the horizon remains at $x=1$, while the exterior region $x\ll1$
approaches the zero-temperature Lifshitz fixed point. The leading
finite-temperature solution is therefore determined by a boundary-value
problem on $0<x\leq1$: it must approach the Lifshitz solution as
$x\to0$ and remain regular at the horizon $x=1$.

We solve the thermal boundary-value problem using the time normalization
$\chi(1)=0$.
Near the horizon $x=1$, we define the horizon electric field $E_h$ through
$a=-E_h(x-1)+\cdots$. Horizon regularity then gives
\begin{equation}
f(1)=0,
\qquad
a(1)=0,
\qquad
f'_h
=
\frac{E_h^2+2V_h}{4},
\qquad
v_h
=
\frac{V_{,\phi,h}}{f'_h}.
\label{eq:lowT_Lifshitz_horizon}
\end{equation}
The remaining horizon parameters are fixed by shooting toward $x\to0$
and requiring the thermal solution to approach the Lifshitz fixed point
without exciting the modes that diverge in this limit. This gives
\begin{equation}
\phi_h=0.38973,
\qquad
E_h=2.3704,
\qquad
f'_h=-1.6226,
\qquad
C_X=0.17769,
\label{eq:lowT_Lifshitz_thermal_data}
\end{equation}
where
$C_X\equiv\lim_{x\to0}\bigl(\widehat{\chi}-2\log x\bigr)$.

Since $f'(z_h)\propto z_h^{-1}$ and the Lifshitz asymptotics give
$e^{-\chi_h/2}\propto z_h^{-(\mathfrak z-1)}$, one has
$T\propto z_h^{-\mathfrak z}$. For the present fixed point,
$\mathfrak z=2$, and the thermal solution gives
\begin{equation}
T
=
8.4014\,z_h^{-2}.
\label{eq:lowT_Lifshitz_temperature}
\end{equation}

Continuing this limiting thermal solution through the horizon gives two
finite scalar-velocity plateaus,
\begin{equation}
v_{1,L}=1.9419,
\qquad
v_{2,L}=2.0598.
\end{equation}
Their Kasner exponents are therefore
\begin{equation}
p_{t,L}^{(1)}
=
-0.014515,
\qquad
p_{t,L}^{(2)}
=
0.014949.
\label{eq:lowT_Lifshitz_limiting_Kasner}
\end{equation}

The leading finite-temperature correction is controlled by the slowest
irrelevant mode of the zero-temperature Lifshitz solution,
$\lambda_1=0.55721$.
At the thermal scale $z\sim z_h$, its amplitude is
\begin{equation}
\zeta
=
A_s z_h^{-\lambda_1},
\label{eq:lowT_Lifshitz_zeta}
\end{equation}
where $A_s=-0.38848$ is fixed by the zero-temperature domain wall.

To determine how this deformation modifies the interior, we perturb the
limiting thermal solution as
\begin{equation}
Y(x)
=
Y_T(x)
+
\zeta Z(x)
+
\cdots .
\end{equation}
The linear response $Z(x)$ obeys
\begin{equation}
\dot Z
=
M_T(x)Z,
\qquad
M_T(x)
=
\left.
\frac{\partial F}{\partial Y}
\right|_{Y_T(x)},
\end{equation}
with the small-$x$ behavior fixed by the same irrelevant eigenmode,
$Z\sim k_{-\lambda_1}x^{-\lambda_1}$.

Continuing this response through the horizon and into the two kinetic
regions gives
\begin{equation}
{\cal R}_1
\equiv
\frac{d p_t^{(1)}}{d\zeta}
=
3.0933,
\qquad
{\cal R}_2
\equiv
\frac{d p_t^{(2)}}{d\zeta}
=
-3.2810.
\label{eq:lowT_Lifshitz_response}
\end{equation}
The leading low-temperature corrections therefore take the form
\begin{equation}
p_t^{(i)}(T)
=
p_{t,L}^{(i)}
+
c_i T^{\lambda_1/2}
+
o\!\left(T^{\lambda_1/2}\right),
\qquad
i=1,2,
\label{eq:lowT_Lifshitz_Kasner_correction}
\end{equation}
with
\begin{equation}
\frac{\lambda_1}{2}
=
0.27861,
\qquad
c_1=-0.66413,
\qquad
c_2=0.70444.
\label{eq:lowT_Lifshitz_amplitudes}
\end{equation}

\subsection{Running-scalar endpoint}
\label{subsec:lowT_running}

For $\mathcal M_{\rm run}$, the zero-temperature scalar does not approach
a finite IR fixed point but instead runs logarithmically. The asymptotic
solution derived in Appendix~\ref{app:running_endpoint} is
\begin{equation}
\phi^2
=
8\varrho
-
\frac{22}{3}\log\varrho
+
C_\phi
+
O\!\left(
\frac{\log\varrho}{\varrho}
\right),
\qquad
\varrho=\log z.
\label{eq:lowT_running_zeroT_corrected}
\end{equation}
The electric sector is subleading in this region, so the leading
low-temperature geometry is governed by the neutral scalar--gravity
system.

As $T\to0$, the thermal horizon moves deeper into the running region and
the horizon value $P\equiv\phi_h$ becomes large. We therefore use $1/P$
as the expansion parameter and introduce
$\xi\equiv\log(z/z_h)$, with the horizon at $\xi=0$. Writing
\begin{equation}
\phi
=
P+\frac{\psi_1(\xi)}{P}
+\frac{\psi_3(\xi)}{P^3}
+
O(P^{-5}),
\qquad
f
=
P^2F_0(\xi)
+
F_2(\xi)
+
O(P^{-2}),
\end{equation}
the leading equations give
\begin{equation}
F_0(\xi)
=
\frac16
\left(
1-e^{3\xi}
\right),
\qquad
\psi_1(\xi)
=
4\xi.
\label{eq:lowT_running_thermal_expansion}
\end{equation}

The first subleading correction also determines the scalar velocity transmitted
into the interior. At order $P^{-3}$, the scalar equation becomes
\begin{equation}
\ddot{\psi}_3
=
\frac{
3\dot{\psi}_3+48\xi+72
}{
1-e^{3\xi}
}.
\label{eq:lowT_running_psi3}
\end{equation}
Horizon regularity fixes $\dot{\psi}_3(0)=-24$, and the regular solution is
\begin{equation}
\dot{\psi}_3(\xi)
=
-\frac{88}{3}
+
\frac{16\xi}{e^{3\xi}-1}.
\label{eq:lowT_running_psi3_solution}
\end{equation}
As $\xi\to+\infty$, $\dot{\psi}_3\to-88/3$, so that
\begin{equation}
v_1
\equiv
\lim z\phi'
=
\frac4P
-
\frac{88}{3P^3}
+
O(P^{-5}).
\label{eq:lowT_running_velocity}
\end{equation}

It remains to relate the large horizon value $P$ to the physical
temperature. Matching the thermal solution to the zero-temperature
running background \eqref{eq:lowT_running_zeroT_corrected} gives
\begin{equation}
P^2
=
8\varrho_h
-
\frac{22}{3}\log\varrho_h
+
C_\phi
+
O\!\left(
\frac{\log\varrho_h}{\varrho_h}
\right),
\qquad
\varrho_h\equiv\log z_h.
\label{eq:lowT_running_Ph}
\end{equation}
Using the large-$P$ thermal solution
together with the zero-temperature relation
$fe^{-\chi}=c_{\rm IR}^2$ in the overlap region, the Hawking temperature is
\begin{equation}
Tz_h
=
K_T\sqrt{\varrho_h}
\left[
1+
O\left(
\frac{\log\varrho_h}{\varrho_h}
\right)
\right],
\qquad
K_T
=
\frac{\sqrt{3}\,c_{\rm IR}}{2\pi}.
\label{eq:lowT_running_temperature}
\end{equation}

Defining $\ell\equiv\log(1/T)$, Eq.~\eqref{eq:lowT_running_temperature}
implies
\begin{equation}
\varrho_h
=
\ell
+
\frac12\log\ell
+
\log K_T
+
O\!\left(
\frac{\log\ell}{\ell}
\right).
\label{eq:lowT_running_Lh_corrected}
\end{equation}
Substituting this into \eqref{eq:lowT_running_Ph}, we obtain
\begin{equation}
\frac{P^2}{8}
=
\ell
-
\frac5{12}\log\ell
+
\log K_T
+
\frac{C_\phi}{8}
+
O\!\left(
\frac{\log\ell}{\ell}
\right).
\label{eq:lowT_running_log_combination}
\end{equation}

Combining \eqref{eq:lowT_running_velocity} and
\eqref{eq:lowT_running_log_combination} with the Kasner relation
\eqref{eq:lowT_Kasner_map}, and using $v_1v_2=4$ for the subsequent
electric inversion, we obtain
\begin{equation}
\begin{split}
p_t^{(1)}+\frac13
&=
\frac{2/9}{
\ell-\frac{5}{12}\log\ell+c_1^{\rm run}
}
+
O\!\left(\frac{\log\ell}{\ell^2}\right),
\\
1-p_t^{(2)}
&=
\frac{2}{
\ell-\frac{5}{12}\log\ell+c_2^{\rm run}
}
+
O\!\left(\frac{\log\ell}{\ell^2}\right),
\end{split}
\label{eq:lowT_running_main_result_corrected}
\end{equation}
where
\begin{equation}
c_1^{\rm run}=-1.4755,
\qquad
c_2^{\rm run}=-0.14222.
\label{eq:lowT_running_constants}
\end{equation}

\section{Numerical verification of the interior dynamics}
\label{sec:numerical}

We now test the analytic predictions of
Secs.~\ref{sec:interior} and \ref{sec:low_temperature} against the full
nonlinear black-hole solutions. We first present the Kasner evolution
over the full temperature range, and then examine separately the
near-critical and low-temperature asymptotic regimes.

\subsection{Full-temperature evolution}
\label{subsec:num_fullT}

\begin{figure}[h!]
\centering
\includegraphics[width=0.9\textwidth]{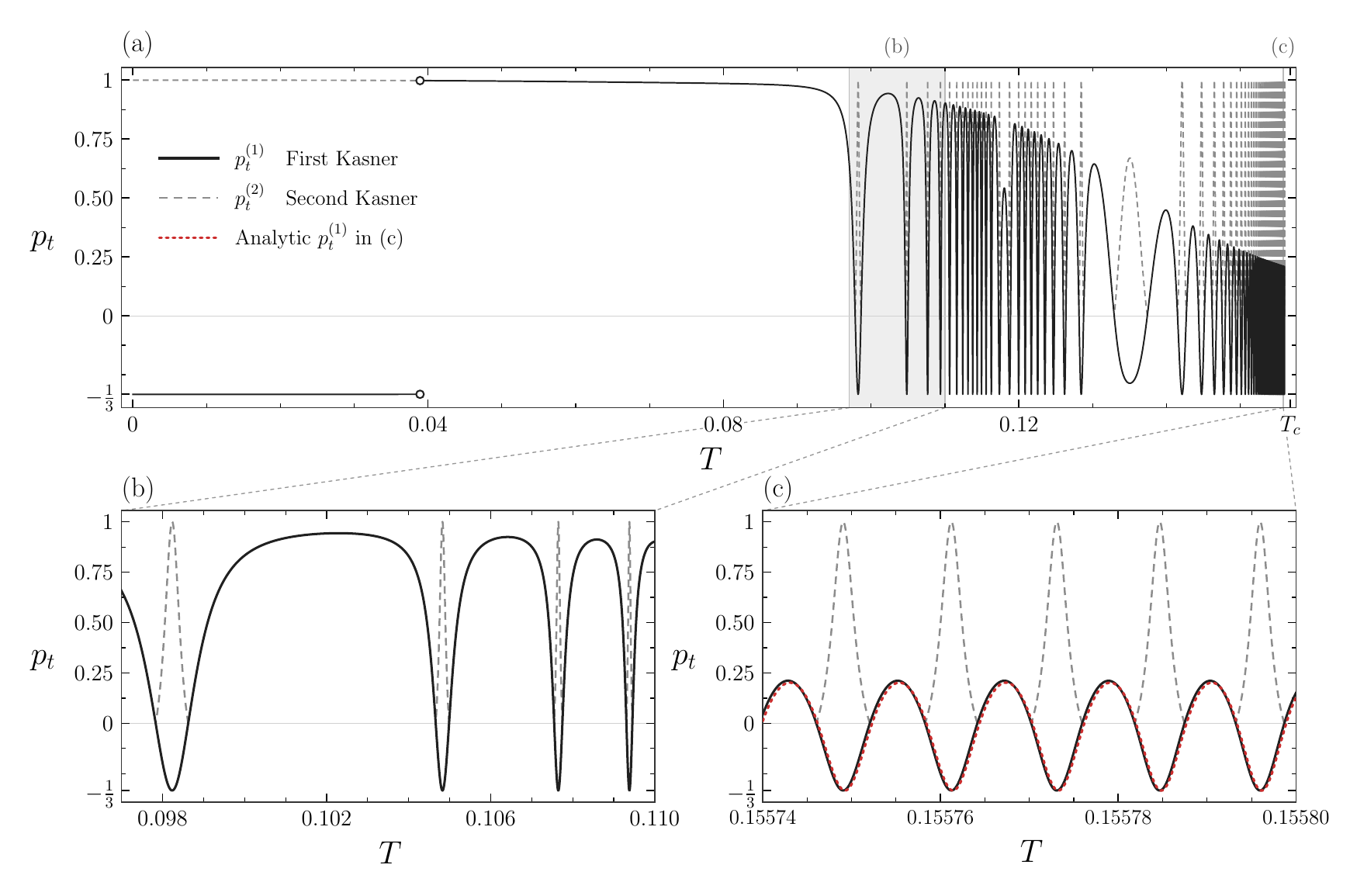}
\caption{\small
Full-temperature evolution of the first- and second-Kasner exponents,
$p_t^{(1)}$ and $p_t^{(2)}$, for $\mathcal M_{\rm AdS}$.
Panel (a) shows the evolution over the full temperature range, with the
black solid and gray dashed curves denoting the first and second Kasner
epochs, respectively. Panel (b) enlarges an intermediate-temperature
window where the Kasner structure undergoes rapid transitions, while
panel (c) probes the asymptotic near-critical regime close to $T_c$.
The red dotted curve in panel (c) denotes the analytic prediction for
$p_t^{(1)}$ from Eq.~\eqref{eq:sec4_explicit_waveform}.
}
\label{fig:num_AdS_Kasner}
\end{figure}

\begin{figure}[h!]
\centering
\includegraphics[width=0.9\textwidth]{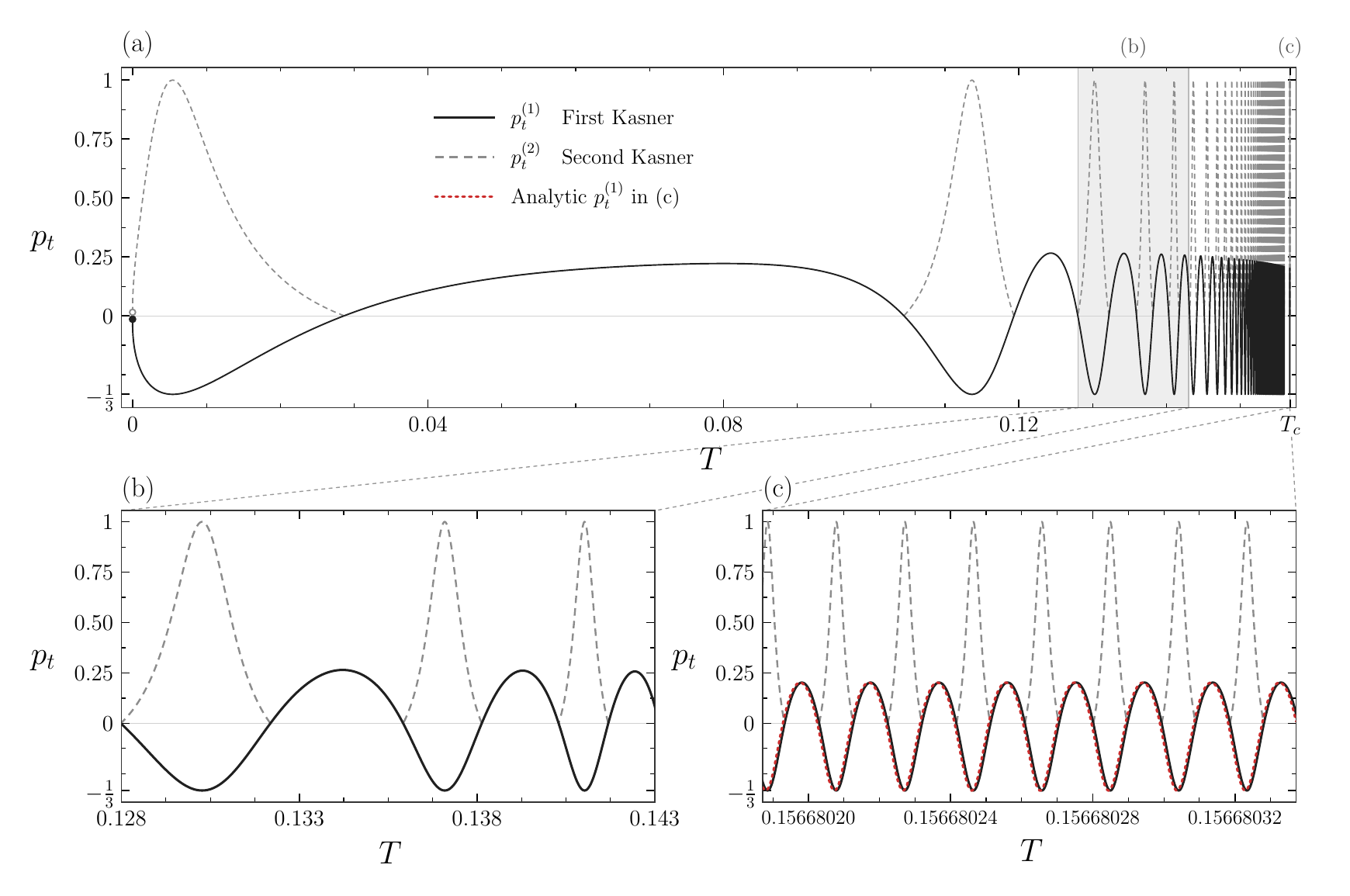}
\caption{\small
Full-temperature evolution of the first- and second-Kasner exponents,
$p_t^{(1)}$ and $p_t^{(2)}$, for $\mathcal M_{\rm Lif}$.
Panel (a) shows the evolution over the full temperature range, with the
black solid and gray dashed curves denoting the first and second Kasner
epochs, respectively. Panel (b) enlarges an intermediate-temperature
window where the Kasner structure undergoes rapid transitions, while
panel (c) probes the asymptotic near-critical regime close to $T_c$.
The red dotted curve in panel (c) denotes the analytic prediction for
$p_t^{(1)}$ from Eq.~\eqref{eq:sec4_explicit_waveform}.
}
\label{fig:num_Lif_Kasner}
\end{figure}

\begin{figure}[h!]
\centering
\includegraphics[width=0.9\textwidth]{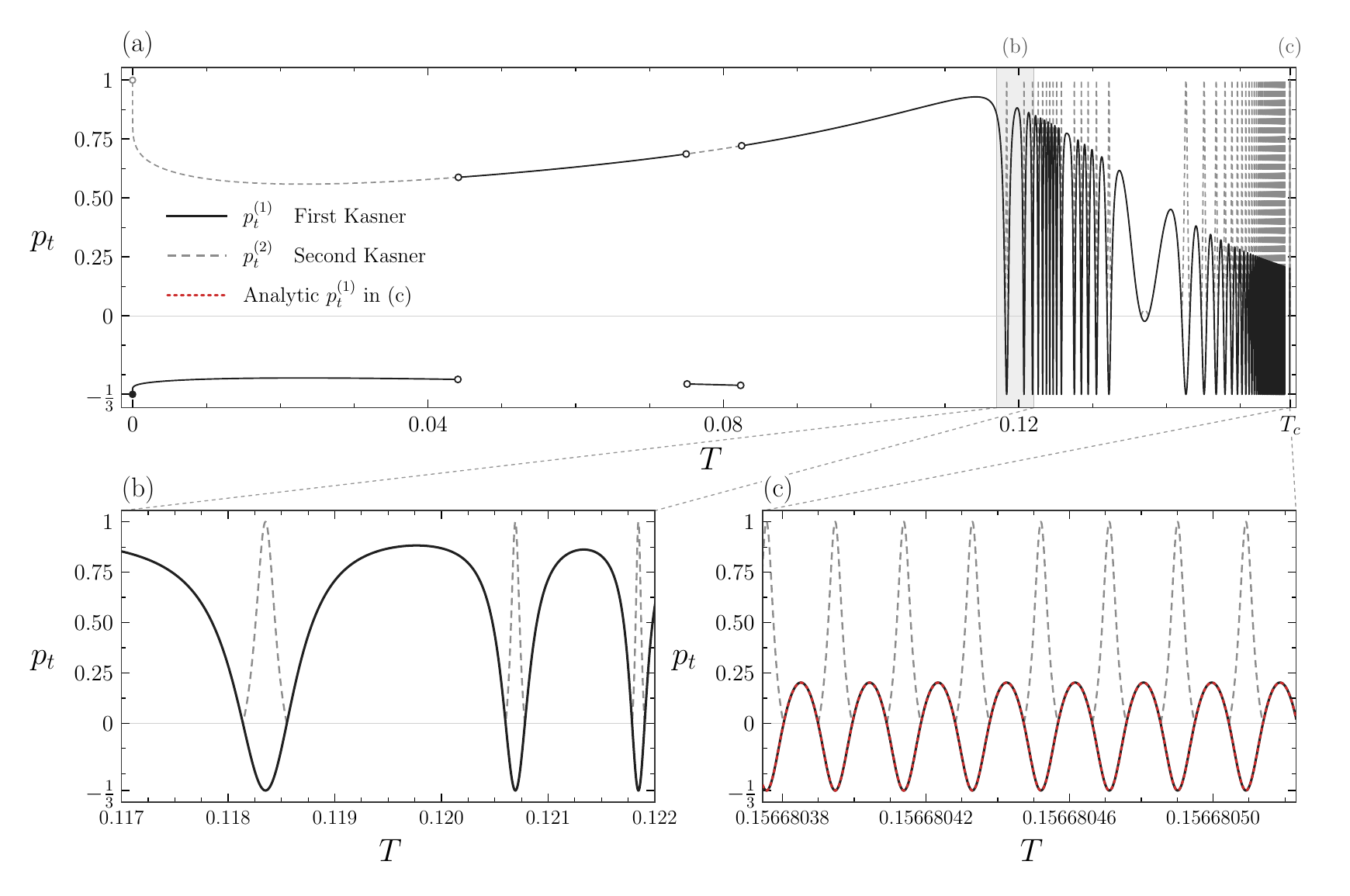}
\caption{\small
Full-temperature evolution of the first- and second-Kasner exponents,
$p_t^{(1)}$ and $p_t^{(2)}$, for $\mathcal M_{\rm run}$.
Panel (a) shows the evolution over the full temperature range, with the
black solid and gray dashed curves denoting the first and second Kasner
epochs, respectively. Panel (b) enlarges an intermediate-temperature
window where the Kasner structure undergoes rapid transitions, while
panel (c) probes the asymptotic near-critical regime close to $T_c$.
The red dotted curve in panel (c) denotes the analytic prediction for
$p_t^{(1)}$ from Eq.~\eqref{eq:sec4_explicit_waveform}.
}
\label{fig:num_Run_Kasner}
\end{figure}

\begin{figure}[h!]
\centering
\includegraphics[width=0.9\textwidth]{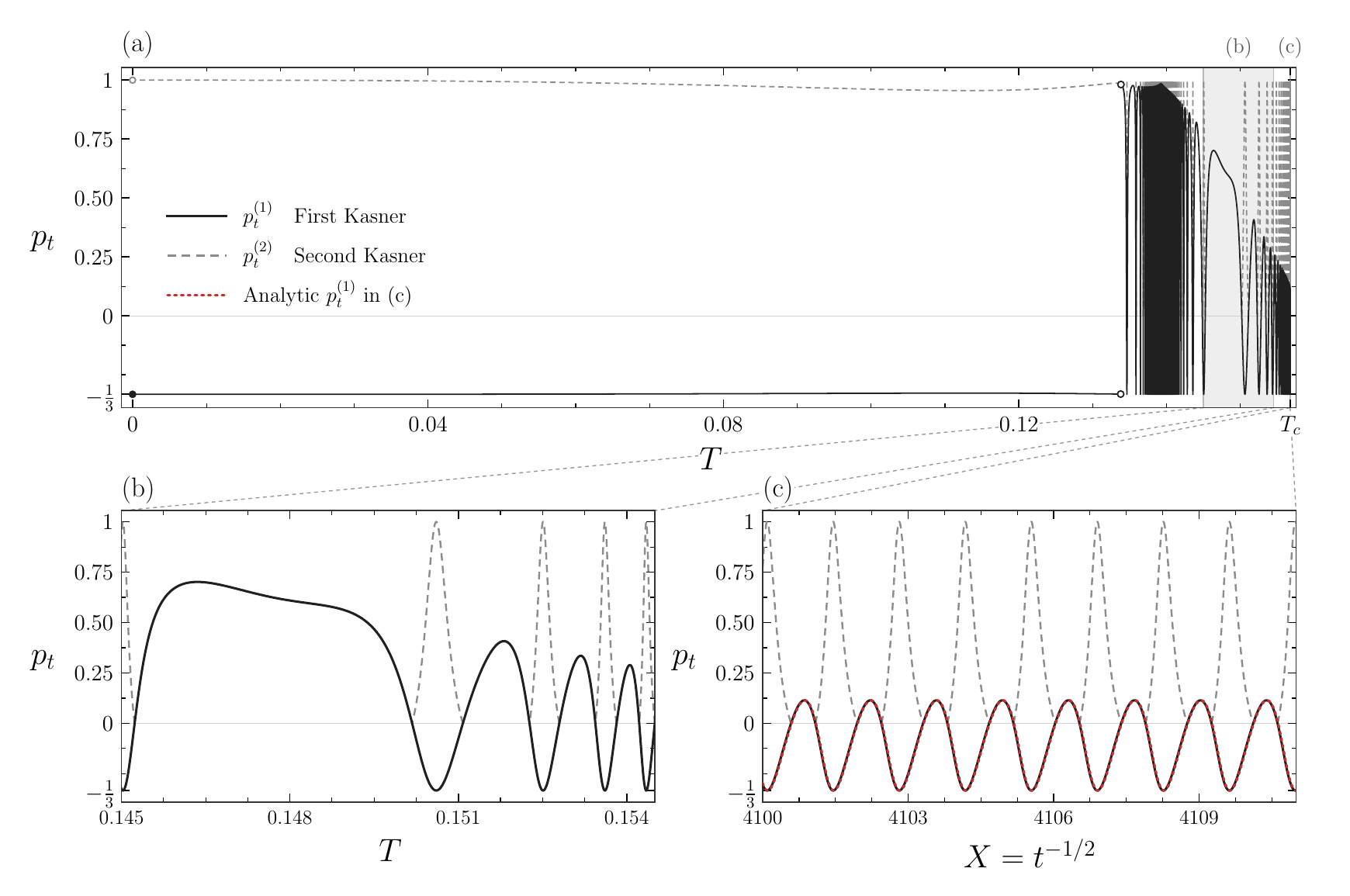}
\caption{\small
Full-temperature evolution of the first- and second-Kasner exponents,
$p_t^{(1)}$ and $p_t^{(2)}$, for the tricritical point.
Panel (a) shows the evolution over the full temperature range, with the
black solid and gray dashed curves denoting the first and second Kasner
epochs, respectively. Panel (b) enlarges an intermediate-temperature
window where the Kasner structure undergoes rapid transitions, while
panel (c) probes the asymptotic near-critical regime close to $T_c$.
The red dotted curve in panel (c) denotes the analytic prediction for
$p_t^{(1)}$ from Eq.~\eqref{eq:tricritical_Kasner_waveform}.
}
\label{fig:num_tri_Kasner}
\end{figure}

Figs.~\ref{fig:num_AdS_Kasner}--\ref{fig:num_tri_Kasner} show the
temperature evolution of the first two Kasner epochs for the four cases.
In all cases, the oscillations become increasingly rapid as $T\to T_c$,
while their behavior away from criticality depends strongly on the
nonlinear completion. The enlarged panels resolve the intermediate and
asymptotic near-critical regimes.

At some intermediate temperatures, the earliest Kasner plateau becomes
too short to be identified reliably. The first visible plateau in the
numerical solution may then correspond to the next Kasner epoch.
Consequently, the plotted first-epoch curve can terminate and reappear
on a different branch as the temperature is varied. This is only a
change in which Kasner plateau is being resolved numerically, not a
discontinuity of the black-hole solution or an additional thermodynamic
transition. The second epoch is shown only when a separate later plateau
can be clearly identified.

For $\mathcal M_{\rm AdS}$, Fig.~\ref{fig:num_AdS_Kasner} connects the
near-critical oscillatory regime to the low-temperature limits
$p_t^{(1)}\to-1/3$ and $p_t^{(2)}\to1$.
The two low-temperature branches correspond to the first and second
Kasner epochs, respectively, and are related by the electric Kasner
inversion in Eq.~\eqref{eq:electric_Kasner_inversion}.

For $\mathcal M_{\rm Lif}$, Fig.~\ref{fig:num_Lif_accumulation} shows a
qualitatively different low-temperature limit: both Kasner exponents
approach finite non-Schwarzschild values. Near $T_c$, however, the
first-Kasner oscillations approach the same leading waveform as in
$\mathcal M_{\rm AdS}$, reflecting the common critical RN background
and transmitted zero mode.

For $\mathcal M_{\rm run}$, Fig.~\ref{fig:num_Run_Kasner} shows that the
Kasner exponents approach the same limiting values as in
$\mathcal M_{\rm AdS}$ but much
more slowly. As shown below, this slow convergence is
inverse logarithmic rather than a power law.

The tricritical case is shown in
Fig.~\ref{fig:num_tri_Kasner}. Near $T_c$, the Kasner oscillations again
accumulate toward the critical point, but with the tricritical scaling
$\Theta_J\sim t^{-1/2}$ and the logarithmic phase correction derived in
Sec.~\ref{subsec:Kasner_scaling}. This gives a denser accumulation than
at an ordinary critical point, with $t_n\sim n^{-2}$.

The full-temperature results provide a global view of these different
regimes. The near-critical ER--Josephson--Kasner description applies
as $T\to T_c$, whereas the low-temperature behavior is controlled
instead by the infrared geometries analyzed in
Sec.~\ref{sec:low_temperature}.

\subsection{Near-critical scaling}
\label{subsec:num_critical}

The full-temperature evolution discussed above provides a global view of the interior dynamics. We now focus on the asymptotic regime $T\to T_c$, where the analytic results of Sec.~\ref{sec:interior} give quantitative predictions for the ER collapse, Josephson phase, and first-Kasner behavior. We consider the ordinary and tricritical cases separately, since their different exterior critical scaling leads to distinct near-critical interior laws.

\subsubsection{Ordinary critical points}

\begin{figure}[!htbp]
\centering
\includegraphics[width=0.98\textwidth]{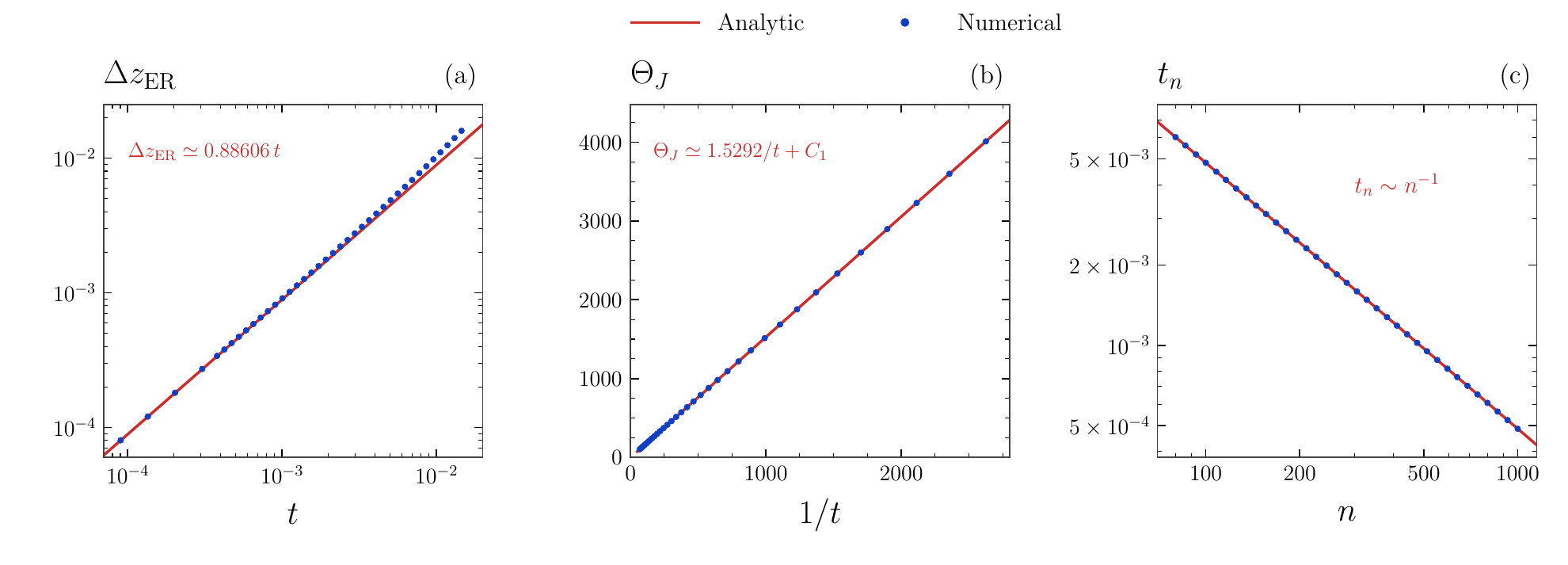}
\caption{\small
Near-critical scaling for $\mathcal M_{\rm AdS}$.
Panels (a)--(c) show the Einstein--Rosen collapse width
$\Delta z_{\rm ER}$, the Josephson phase $\Theta_J$, and the
accumulation sequence $t_n$ of the first-Kasner scalar-velocity zeros,
respectively. Blue points are full nonlinear results, while the red
curves denote the analytic predictions in
\eqref{eq:temperature_ER_widthnum} and
\eqref{eq:Kasner_zero_accumulation2nd}.
}
\label{fig:num_AdS_er}
\end{figure}

\begin{figure}[!htbp]
\centering
\includegraphics[width=0.98\textwidth]{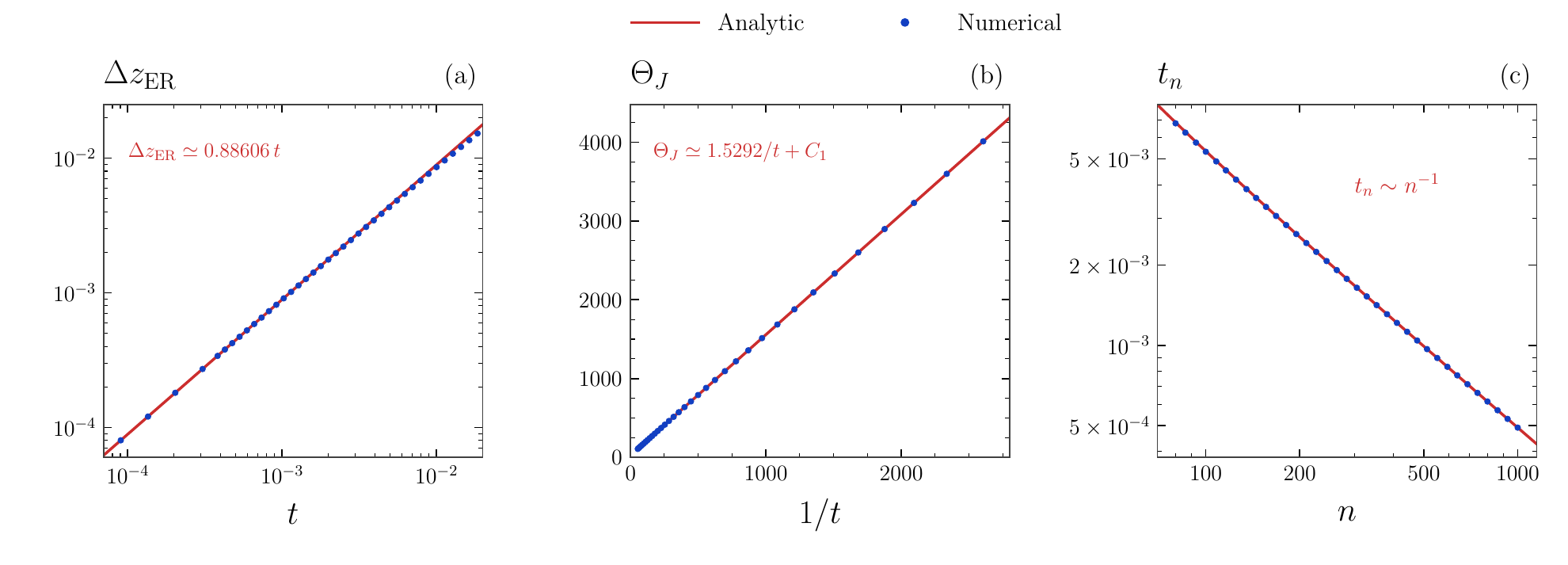}
\caption{\small
Near-critical scaling for $\mathcal M_{\rm Lif}$.
Panels (a)--(c) show the Einstein--Rosen collapse width
$\Delta z_{\rm ER}$, the Josephson phase $\Theta_J$, and the
accumulation sequence $t_n$ of the first-Kasner scalar-velocity zeros,
respectively. Blue points are full nonlinear results, while the red
curves denote the analytic predictions in
\eqref{eq:temperature_ER_widthnum} and
\eqref{eq:Kasner_zero_accumulation2nd}.
}
\label{fig:num_Lif_er}
\end{figure}

\begin{figure}[!htbp]
\centering
\includegraphics[width=0.98\textwidth]{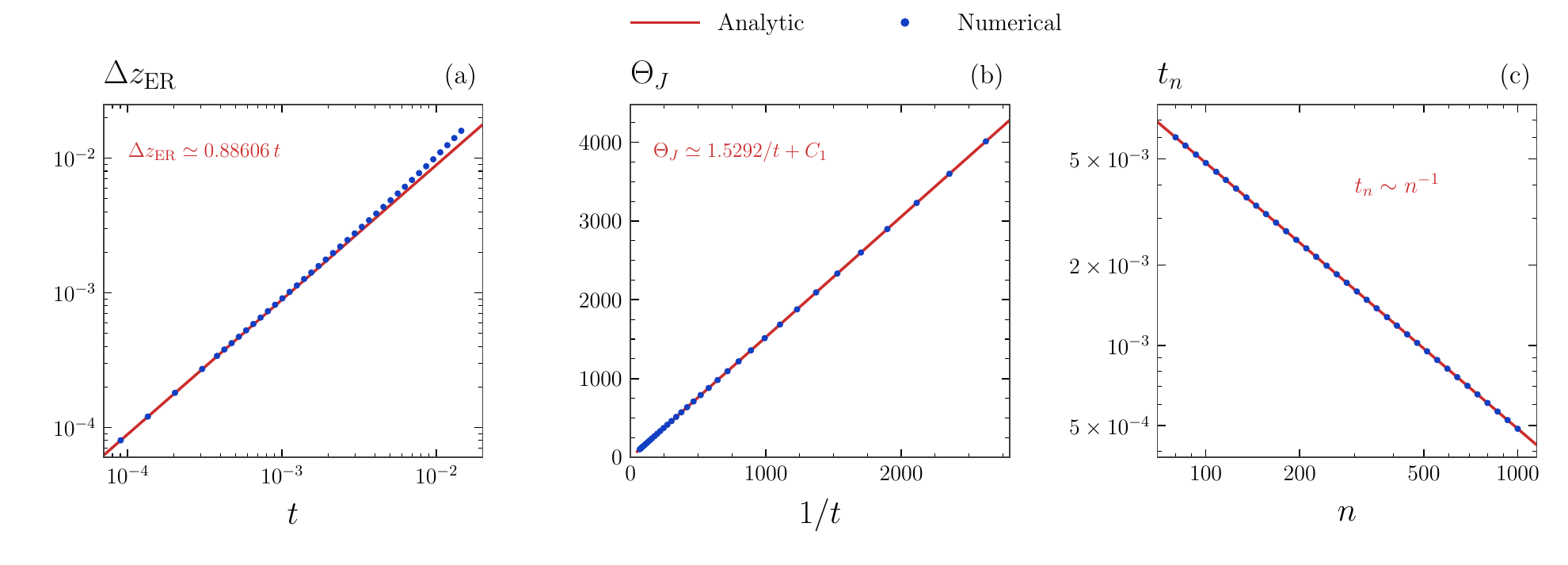}
\caption{\small
Near-critical scaling for $\mathcal M_{\rm run}$.
Panels (a)--(c) show the Einstein--Rosen collapse width
$\Delta z_{\rm ER}$, the Josephson phase $\Theta_J$, and the
accumulation sequence $t_n$ of the first-Kasner scalar-velocity zeros,
respectively. Blue points are full nonlinear results, while the red
curves denote the analytic predictions in
\eqref{eq:temperature_ER_widthnum} and
\eqref{eq:Kasner_zero_accumulation2nd}.
}
\label{fig:num_run_er}
\end{figure}

The three ordinary critical models share the same critical RN geometry,
zero-mode transmission amplitude, and leading temperature coefficient.
They therefore obey the common collapse and phase scaling in
Eq.~\eqref{eq:temperature_ER_widthnum}, together with the accumulation
law in Eq.~\eqref{eq:Kasner_zero_accumulation2nd}.
As shown in Figs.~\ref{fig:num_AdS_er}, \ref{fig:num_Lif_er} and
\ref{fig:num_run_er}, the numerical solutions approach these analytic
predictions as $t\to0$. The finite phase offset $C_1$ is determined
separately for each nonlinear completion and kept fixed in the
subsequent comparisons.

Keeping the finite phase offset, the ordinary-critical zero sequence is
\begin{equation}
t_n
\simeq
\frac{B_1}{\pi n-C_1},
\qquad
\frac{B_1}{\pi}
=
0.48677.
\label{eq:ordinary_zero_sequence}
\end{equation}
Independent fits to the three ordinary critical models give
\begin{align}
\mathcal M_{\rm AdS}: \qquad
t_n
&\simeq
0.48672\,(n+0.68544)^{-0.99999},
\nonumber\\
\mathcal M_{\rm Lif}: \qquad
t_n
&\simeq
0.48874\,(n-8.4530)^{-1.0005},
\nonumber\\
\mathcal M_{\rm run}: \qquad
t_n
&\simeq
0.48674\,(n+0.80762)^{-0.99999}.
\label{eq:ordinary_zero_fits}
\end{align}
The fitted exponents are all consistent with the predicted
$t_n\sim n^{-1}$ accumulation, while the leading coefficients agree
closely with $B_1/\pi=0.48677$. The model-dependent shifts encode the
finite phase offsets together with subleading finite-$n$ corrections.
The red curves in Figs.~\ref{fig:num_AdS_er}, \ref{fig:num_Lif_er} and
\ref{fig:num_run_er} use the analytic relation
Eq.~\eqref{eq:ordinary_zero_sequence}.
Here $t_n$ denotes the zeros of the first-Kasner scalar velocity,
$v_1(t_n)=0$.

\begin{figure}[!ht]
\centering
\includegraphics[width=0.95\textwidth]{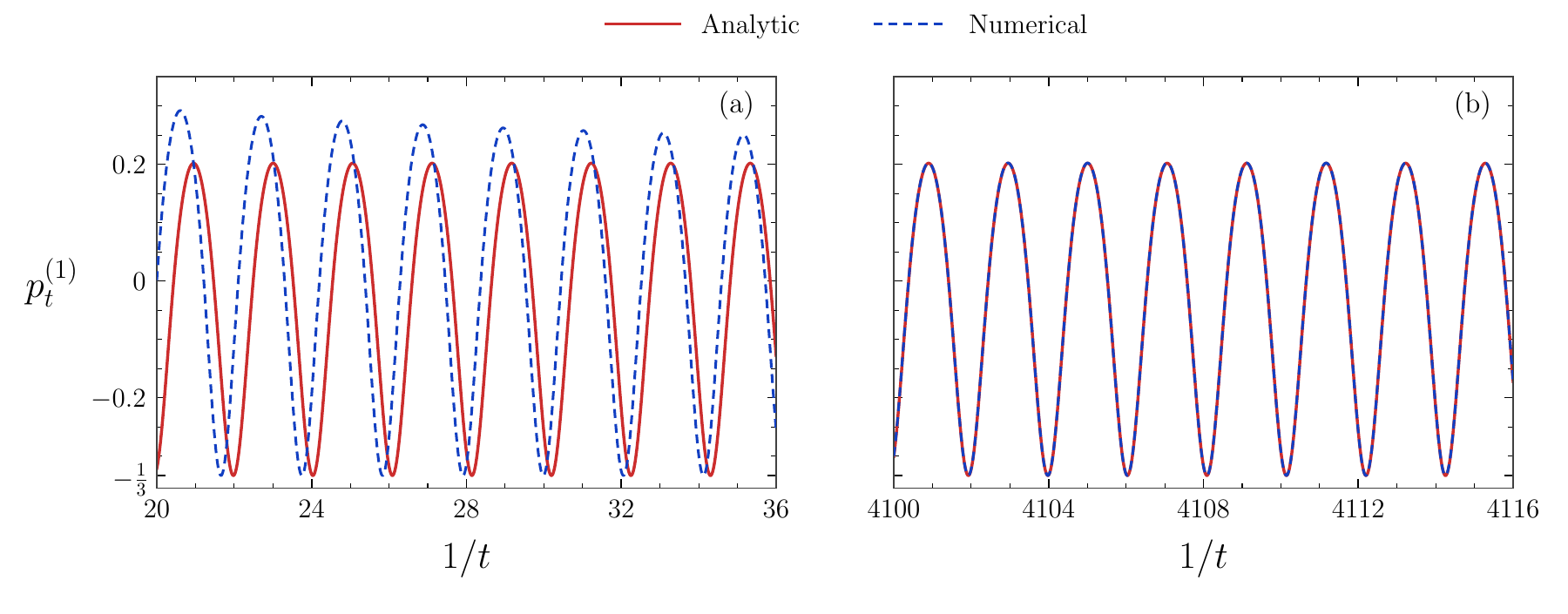}
\caption{\small
Near-critical first-Kasner exponent $p_t^{(1)}$ for
$\mathcal M_{\rm AdS}$ as a function of $1/t$.
Panel (a) shows an intermediate near-critical window, where subleading
corrections remain visible, while panel (b) probes the asymptotic regime
closer to $T_c$.
The blue dashed curves are obtained from the full nonlinear solutions,
and the red solid curves denote the analytic prediction
Eq.~\eqref{eq:sec4_explicit_waveform}.
}
\label{fig:num_AdS_accumulation}
\end{figure}

\begin{figure}[!ht]
\centering
\includegraphics[width=0.95\textwidth]{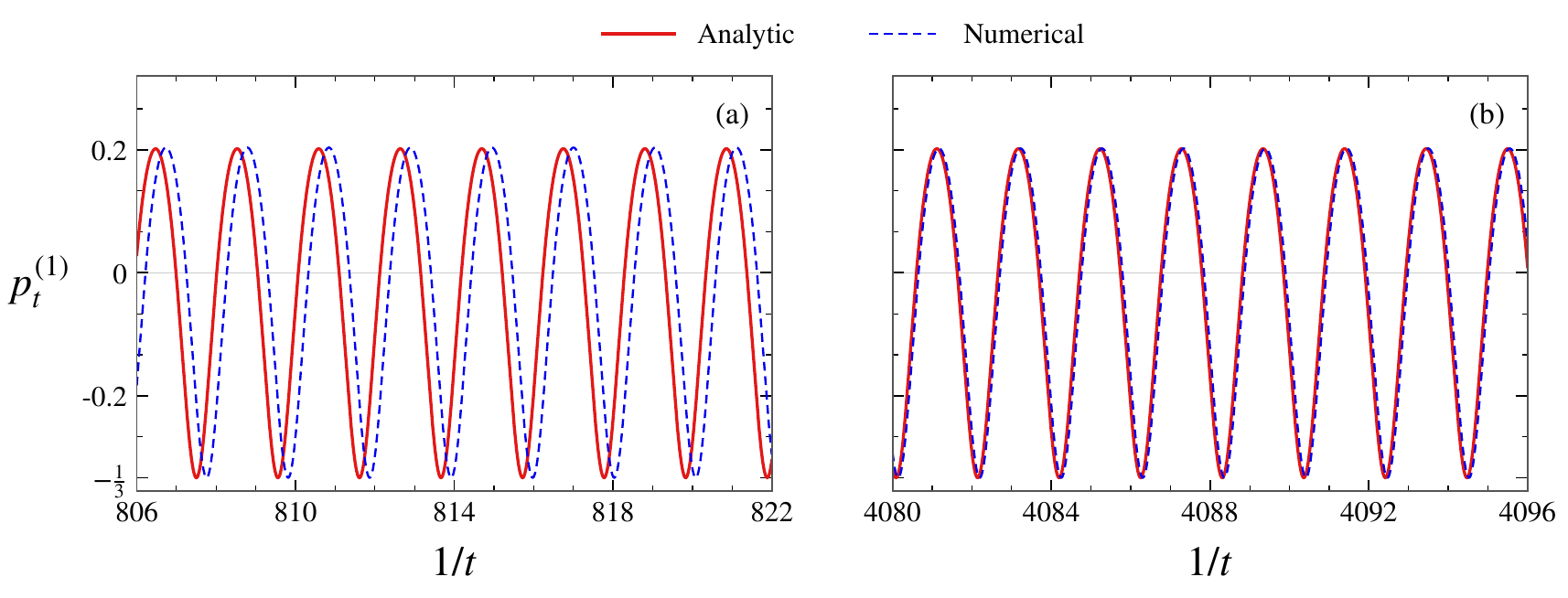}
\caption{\small
Near-critical first-Kasner exponent $p_t^{(1)}$ for
$\mathcal M_{\rm Lif}$ as a function of $1/t$.
Panel (a) shows an intermediate near-critical window, where subleading
corrections remain visible, while panel (b) probes the asymptotic regime
closer to $T_c$.
The blue dashed curves are obtained from the full nonlinear solutions,
and the red solid curves denote the analytic prediction
Eq.~\eqref{eq:sec4_explicit_waveform}.
}
\label{fig:num_Lif_accumulation}
\end{figure}

\begin{figure}[!ht]
\centering
\includegraphics[width=0.95\textwidth]{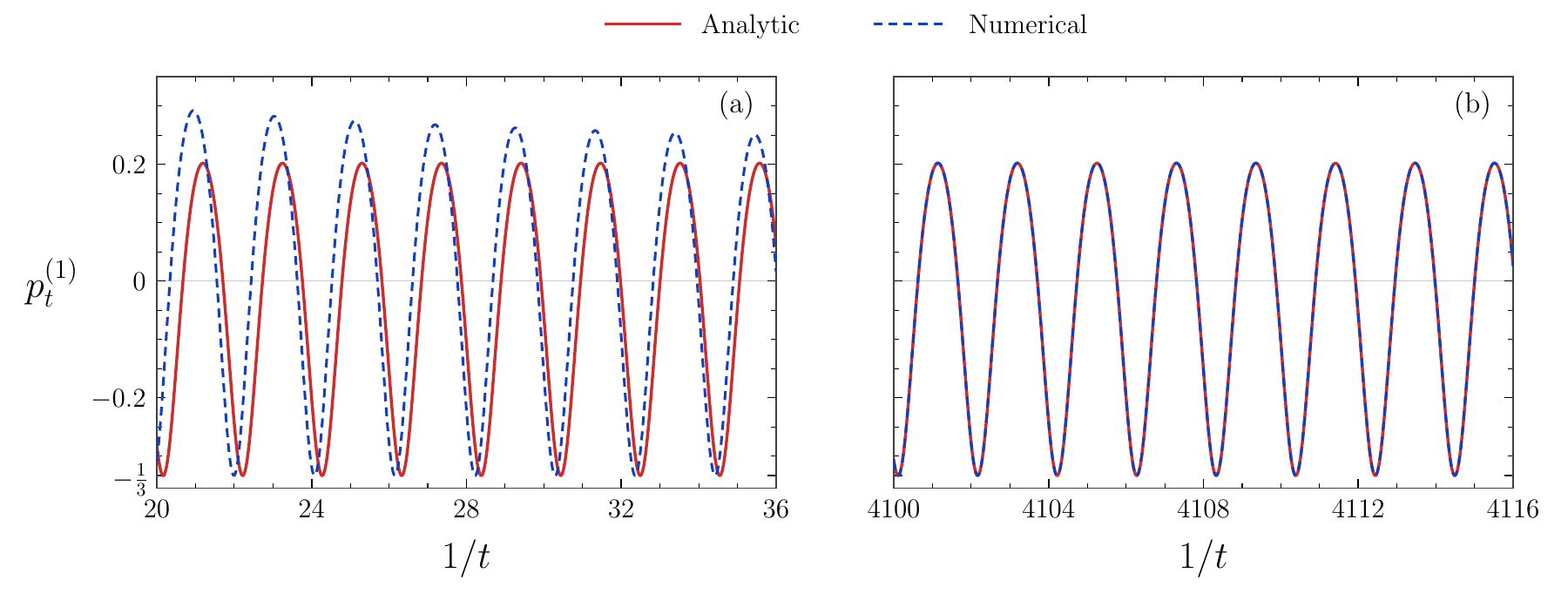}
\caption{\small
Near-critical first-Kasner exponent $p_t^{(1)}$ for
$\mathcal M_{\rm run}$ as a function of $1/t$.
Panel (a) shows an intermediate near-critical window, where subleading
corrections remain visible, while panel (b) probes the asymptotic regime
closer to $T_c$.
The blue dashed curves are obtained from the full nonlinear solutions,
and the red solid curves denote the analytic prediction
Eq.~\eqref{eq:sec4_explicit_waveform}.
}
\label{fig:num_Run_accumulation}
\end{figure}

The waveform comparisons in
Figs.~\ref{fig:num_AdS_accumulation},
\ref{fig:num_Lif_accumulation} and
\ref{fig:num_Run_accumulation}
provide a more stringent test of the near-critical prediction.
Finite-$t$ deviations are visible in the intermediate windows, but are
strongly suppressed closer to $T_c$, where the numerical curves approach
the analytic waveform. This agreement tests not only the Josephson phase
accumulation but also its transfer to the first Kasner epoch.

\subsubsection{Tricritical point}

\begin{figure}[!htbp]
\centering
\includegraphics[width=0.98\textwidth]{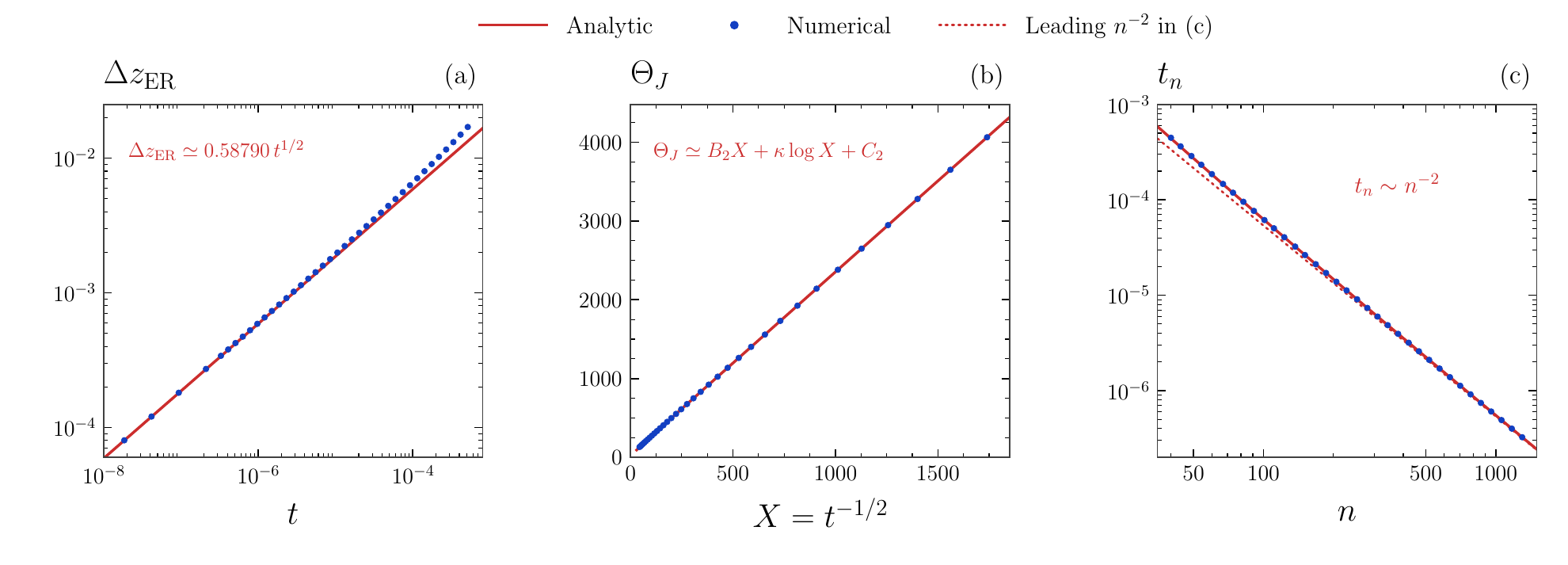}
\caption{\small
Near-critical scaling for the tricritical point.
Panels (a)--(c) show the Einstein--Rosen collapse width
$\Delta z_{\rm ER}$, the Josephson phase $\Theta_J$, and the
accumulation sequence $t_n$ of the first-Kasner scalar-velocity zeros,
respectively. Blue points are full nonlinear results, while the red
curves denote the analytic predictions in
\eqref{eq:temperature_ER_widthnumtri} and
\eqref{eq:tricritical_accumulation}.
}
\label{fig:num_tri_er}
\end{figure}

We next consider the tricritical point introduced in
Sec.~\ref{subsec:sc_tricritical}. The corresponding ER-collapse width and
Josephson phase are given analytically by
Eq.~\eqref{eq:temperature_ER_widthnumtri}, including the logarithmic
phase correction. As shown in Fig.~\ref{fig:num_tri_er},
the full nonlinear solutions reproduce both the square-root collapse
scaling and the logarithmically corrected phase as $t\to0$. The finite
phase offset extracted from the numerical solution is
$C_2=0.78273$ in the convention used in the figure.

The nonlinear Josephson-to-Kasner transfer can also be extracted directly
from the tricritical solutions. Over one period, the numerical transfer
function is accurately represented by the odd-harmonic expansion
\begin{align}
{\cal F}(\Theta)
\simeq{}&
-1.11664451\cos\Theta
-0.44667042\sin\Theta
\nonumber\\
&+0.05164517\cos3\Theta
-0.04142294\sin3\Theta
\nonumber\\
&+0.00043649\cos5\Theta
+0.00282265\sin5\Theta .
\label{eq:num_tricritical_transfer}
\end{align}
The appearance of only odd harmonics is consistent with the
antiperiodicity
${\cal F}(\Theta+\pi)=-{\cal F}(\Theta)$.

The relevant zero of Eq.~\eqref{eq:num_tricritical_transfer} fixes the
phase entering the accumulation condition
Eq.~\eqref{eq:Kasner_zero_phase_condition}. Together with the finite
phase offset $C_2$, this determines the tricritical zero sequence.
The resulting sequence is shown in Fig.~\ref{fig:num_tri_er} and
approaches $t_n\simeq0.53824\,n^{-2}$ at large $n$, in agreement with
Eq.~\eqref{eq:tricritical_accumulation}. The logarithmic phase correction
remains important for the finite values of $n$ shown in the figure.

\begin{figure}[!ht]
\centering
\includegraphics[width=0.95\textwidth]{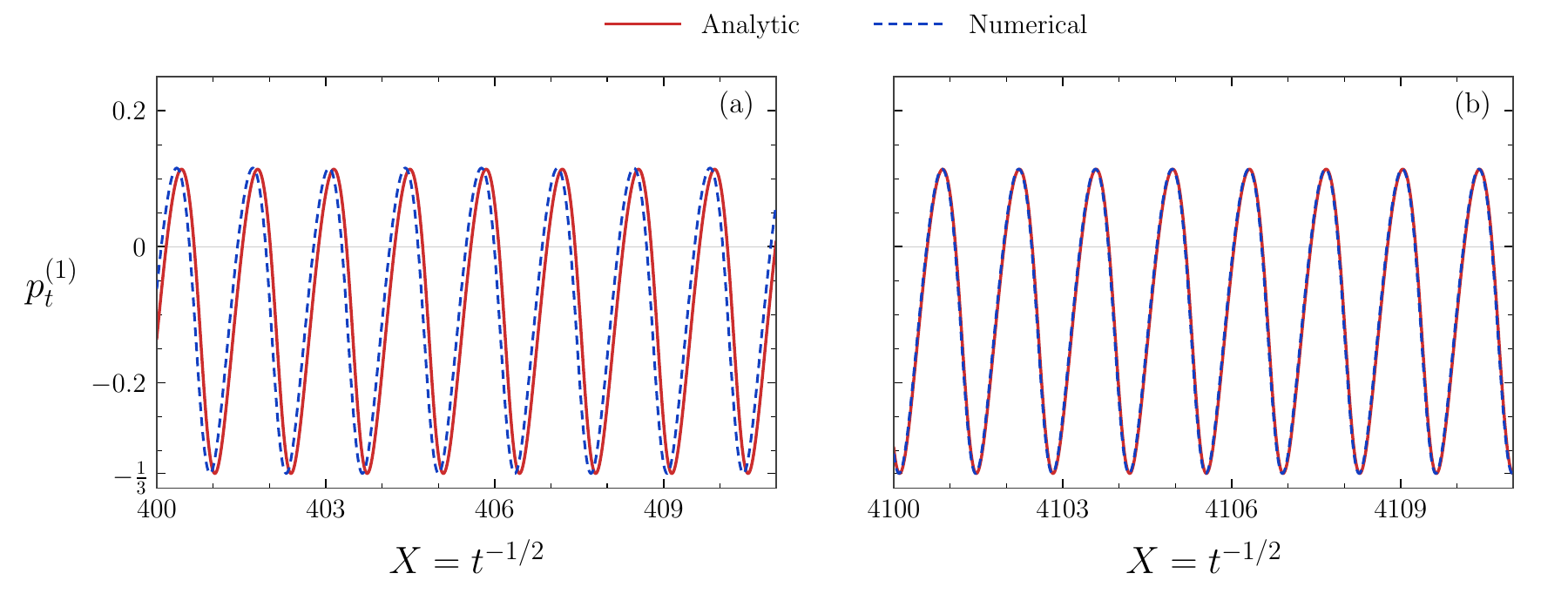}
\caption{\small
Near-critical first-Kasner exponent $p_t^{(1)}$ for
the tricritical point as a function of $t^{-1/2}$.
Panel (a) shows an intermediate near-critical window, where subleading
corrections remain visible, while panel (b) probes the asymptotic regime
closer to $T_c$.
The blue dashed curves are obtained from the full nonlinear solutions,
and the red solid curves denote the analytic prediction
Eq.~\eqref{eq:tricritical_Kasner_waveform}.
}
\label{fig:num_tri_accumulation}
\end{figure}

Fig.~\ref{fig:num_tri_accumulation} provides a direct test of the
nonlinear transfer function. Combining
Eq.~\eqref{eq:num_tricritical_transfer} with the Josephson phase in
Eq.~\eqref{eq:temperature_ER_widthnumtri} gives the analytic
first-Kasner waveform through Eq.~\eqref{eq:tricritical_Kasner_waveform}.
The agreement improves as the critical point is approached, while
finite-$t$ corrections remain visible in the intermediate window.

\subsection{Low-temperature scaling}
\label{subsec:num_lowT}

The three infrared completions lead to qualitatively different
low-temperature Kasner behavior. We now compare the full nonlinear
solutions with the analytic predictions of
Sec.~\ref{sec:low_temperature} for the cohesive AdS$_4$, Lifshitz, and
running-scalar endpoints.

\subsubsection{AdS$_4$ endpoint}

\begin{figure}[!htbp]
\centering
\includegraphics[width=0.55\textwidth]{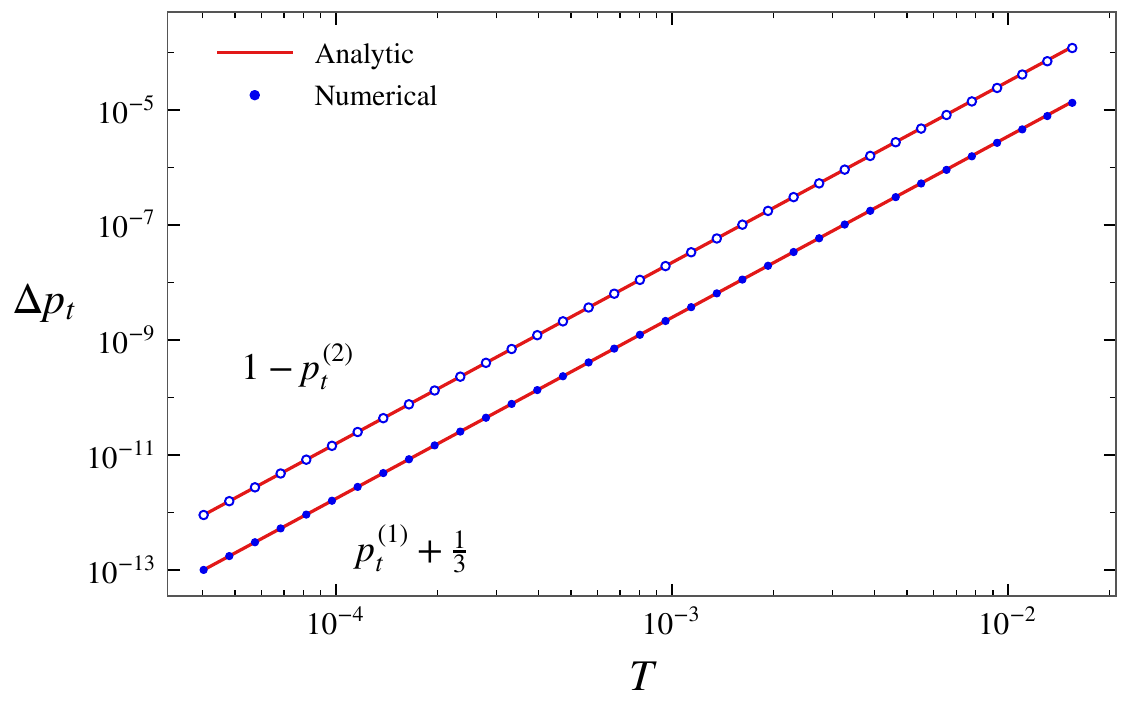}
\caption{\small
Low-temperature deviations of the first- and second-Kasner exponents for
$\mathcal M_{\rm AdS}$.
Filled blue circles show $p_t^{(1)}+1/3$, while open blue circles show
$1-p_t^{(2)}$.
The red lines denote the analytic low-temperature predictions in
Eq.~\eqref{eq:lowT_AdS4_main_result}.
}
\label{fig:num_AdS_IR}
\end{figure}

For $\mathcal M_{\rm AdS}$, the numerical fits give
\begin{equation}
\begin{split}
p_t^{(1)}+\frac13
&\simeq
6.8576\,T^{3.1481},
\qquad
1-p_t^{(2)}
\simeq
61.719\,T^{3.1481}.
\end{split}
\label{eq:num_AdS_lowT_fit}
\end{equation}
These fits agree closely with the analytic prediction
Eq.~\eqref{eq:lowT_AdS4_main_result}, which gives the common exponent
$2\delta_\phi=3.1482$ and coefficients $6.8623$ and $61.760$.
The fitted amplitudes differ from the analytic values by less than
$0.07\%$, while their ratio approaches $9$, as implied by the electric
Kasner inversion. Figure~\ref{fig:num_AdS_IR} shows the resulting
agreement over the low-temperature regime.

\subsubsection{Lifshitz endpoint}

\begin{figure}[!htbp]
\centering
\includegraphics[width=0.9\textwidth]{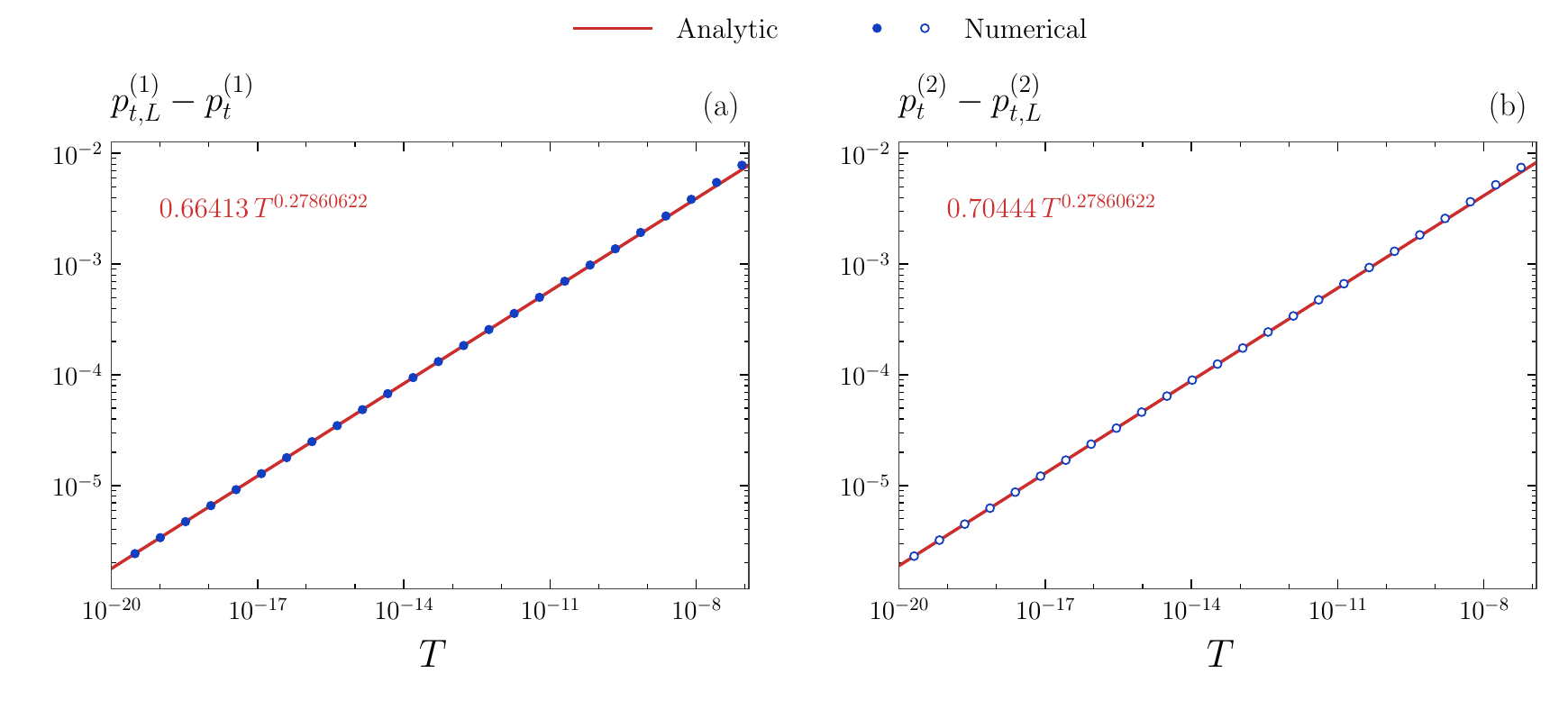}
\caption{\small
Low-temperature deviations from the limiting Kasner exponents for
$\mathcal M_{\rm Lif}$.
Panel (a) shows $p_{t,L}^{(1)}-p_t^{(1)}$, while panel (b) shows
$p_t^{(2)}-p_{t,L}^{(2)}$.
Filled and open blue circles are the full nonlinear results, and the red
lines denote the analytic low-temperature predictions in
Eqs.~\eqref{eq:lowT_Lifshitz_Kasner_correction} and
\eqref{eq:lowT_Lifshitz_amplitudes}.
}
\label{fig:num_Lif_IR}
\end{figure}

For $\mathcal M_{\rm Lif}$, we measure the approach to the limiting
Kasner exponents in Eq.~\eqref{eq:lowT_Lifshitz_limiting_Kasner} through
the positive deviations
\begin{equation}
\Delta p_t^{(1)}
=
p_{t,L}^{(1)}-p_t^{(1)},
\qquad
\Delta p_t^{(2)}
=
p_t^{(2)}-p_{t,L}^{(2)}.
\end{equation}
The numerical fits give
\begin{equation}
\Delta p_t^{(1)}
\simeq
0.66500\,T^{0.27864},
\qquad
\Delta p_t^{(2)}
\simeq
0.70550\,T^{0.27864}.
\end{equation}
The fitted exponents agree with the analytic value
$\lambda_1/2=0.27861$, while the amplitudes agree with the predictions
$c_1=0.66413$ and $c_2=0.70444$ in
Eqs.~\eqref{eq:lowT_Lifshitz_Kasner_correction} and
\eqref{eq:lowT_Lifshitz_amplitudes} to within $0.16\%$.
Figure~\ref{fig:num_Lif_IR} shows the resulting agreement over the
low-temperature regime.

\subsubsection{Running-scalar endpoint}

\begin{figure}[!htbp]
\centering
\includegraphics[width=0.55\textwidth]{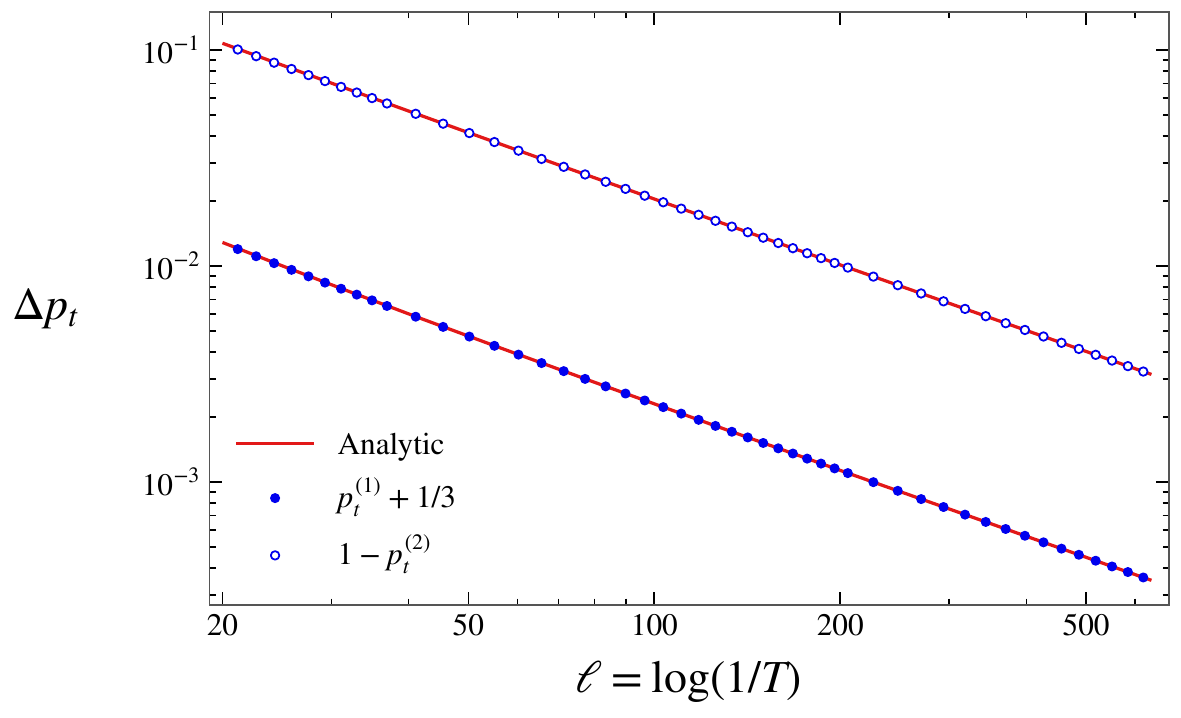}
\caption{\small
Low-temperature deviations of the first- and second-Kasner exponents for
$\mathcal M_{\rm run}$, plotted against
$\ell=\log(1/T)$.
Filled blue circles show $p_t^{(1)}+1/3$, while open blue circles show
$1-p_t^{(2)}$.
The red curves denote the analytic low-temperature prediction in
Eq.~\eqref{eq:lowT_running_main_result_corrected}.
}
\label{fig:num_Run_IR}
\end{figure}

For $\mathcal M_{\rm run}$, the approach to
$p_t^{(1)}=-1/3$ and $p_t^{(2)}=1$ is inverse logarithmic.
With $\ell=\log(1/T)$, direct power-law fits give
\begin{equation}
p_t^{(1)}+\frac13
\simeq
0.23848\,\ell^{-1.0101},
\qquad
1-p_t^{(2)}
\simeq
2.0894\,\ell^{-1.0062}.
\label{eq:num_running_lowT_fit}
\end{equation}
The fitted powers are close to the leading $\ell^{-1}$ behavior, while
the residual drift is accounted for by the subleading logarithmic terms
in Eq.~\eqref{eq:lowT_running_main_result_corrected}. As shown in
Fig.~\ref{fig:num_Run_IR}, the full numerical solutions approach this
analytic prediction over the low-temperature regime, with agreement at
the $10^{-5}$ level at the lowest temperatures reached.

\section{Discussion}
\label{sec:discussion}

In this work, we have shown how the critical behavior of a holographic
superconductor is transmitted quantitatively to the black-hole interior.
Combining the exterior perturbative expansion with the continuation of
the critical zero mode through the reference RN geometry, we find that
the transmitted amplitude and local Cauchy-horizon quantities determine
the leading ER-collapse width and Josephson phase. This extends the
neutral-scalar transmission analysis of Ref.~\cite{Gao:2026CH} to a
charged condensate, for which the Josephson regime introduces an
additional stage between the collapse layer and the first Kasner epoch
\cite{Hartnoll:2020fhc,Zhao:2025nonlinear}.

The resulting critical scaling is controlled by the order at which the
exterior branch departs from the critical solution. If
$t\equiv1-T/T_c\propto\epsilon^{2r}$, the ER-collapse width and Josephson
phase scale as $\Delta z_{\rm ER}\propto t^{1/r}$ and
$\Theta_J\propto t^{-1/r}$, respectively. The rapidly accumulated phase
is then transferred to the first Kasner epoch and, for a simple zero of
the transfer function, produces the accumulation law $t_n\sim n^{-r}$.
Thus ordinary criticality gives $r=1$, whereas the tricritical point has
$r=2$. At tricriticality, the quartic St\"uckelberg interaction further
produces a logarithmic correction to the Josephson phase without changing
its leading power-law divergence. The fully nonlinear solutions confirm
these predictions in both the ordinary and tricritical cases, including
the leading collapse and phase coefficients fixed by the critical zero
mode.

The low-temperature limit reveals a complementary dependence on the
infrared completion. Although the three representative models share the
same critical RN background and hence the same leading near-critical
transmission, they flow to distinct zero-temperature geometries. The
cohesive AdS$_4$ endpoint approaches the limiting Kasner exponents with
a power law set by its leading irrelevant scalar mode, the Lifshitz
endpoint approaches finite non-Schwarzschild Kasner values with a
fractional-power correction governed by its slowest irrelevant mode,
and the running-scalar endpoint converges only inverse logarithmically.
Thus the near-critical interior is controlled by the critical zero-mode
transmission, whereas the low-temperature behavior retains information
about the zero-temperature infrared completion.

The first Kasner epoch does not, in general, determine the final approach
to the singularity. When the electric wall approximation remains valid,
a subsequent epoch is related to the first by the electric Kasner
inversion in Eq.~\eqref{eq:electric_Kasner_inversion}. More general
matter couplings can instead generate additional inversions or repeated
Kasner transitions
\cite{An:2022Mtheory,Henneaux:2022Kasner,Cai:2023igv,
Hartnoll:2022bouncing}. Such later evolution depends on the nonlinear
interior dynamics and is not fixed by the leading critical-mode
transmission alone.

Several directions remain open. Higher multicritical points provide a
natural extension of the scaling structure found here and would test the
general prediction $t_n\sim n^{-r}$ for $r>2$. It would also be useful to
understand the finite matching phases beyond the leading asymptotic
analysis and their dependence on the nonlinear completion. More
fundamentally, an important question is whether the interior scales and
Kasner structure identified here can be reconstructed from observables
defined entirely outside the horizon. Establishing such a connection
would provide a direct probe of how boundary criticality encodes
black-hole interior geometry.

\subsection*{Acknowledgments}
We would like to thank Yan Liu (Beihang University), Ling-Long Gao and Jun-Kun Zhao for useful discussions. This work was supported by the National Natural Science Foundation of China Grants No. 12447134 and 12605122. H-D.~L.~also acknowledges the support from 
the Postdoctoral Innovation Project of Shandong Province SDCX-ZG-202503036. Y. L. is also supported by Natural Science Foundation of Shandong Province under Grants No. ZR2023QA133 and Yantai University under Grants No. WL22B218

\vspace{0.95cm}
\appendix

\section{Boundary conditions and horizon expansions}
\label{app:boundary_horizon}

We collect here the asymptotic data used to construct the black brane solutions. Near the AdS boundary, the fields admit the expansions
\begin{align}
\phi
&=
\phi_1 z+\phi_2 z^2+\cdots,
&
\chi
&=
\chi_0+\frac{\phi_1^2}{4}z^2+\cdots,
\nonumber\\
f
&=
1+\frac{\phi_1^2}{4}z^2+f_3 z^3+\cdots,
&
\Phi
&=
\mu-\rho z
+\frac{q^2\mu\phi_1^2}{2}z^2+\cdots .
\label{app:eq:UV_general}
\end{align}
The residual time rescaling is used to set $\chi_0=0$. Since $m^2=-2$, both standard and alternative quantizations are allowed. Throughout this work we adopt the standard quantization, for which $\phi_1$ is the source and $\phi_2$ determines the condensate. Imposing the source-free condition $\phi_1=0$ gives
\begin{align}
\phi
&=
\phi_2 z^2+\cdots,
&
\Phi
&=
\mu-\rho z+\cdots,
\nonumber\\
f
&=
1+f_3 z^3+\cdots,
&
\chi
&=
\mathcal O(z^4).
\label{app:eq:UV_sourcefree}
\end{align}
We fix the remaining overall scale by setting $\mu=1$, and use
$O\equiv\phi_2$
as the condensate convention in the numerical analysis.


At $z=z_h$, regularity requires $f_h=\Phi_h=0$, where a subscript $h$ denotes evaluation at the horizon, e.g. $G_h\equiv G(\phi_h)$. The first derivatives are then fixed by the equations of motion,
\begin{equation}
\begin{split}
f'_h
&=
\frac{1}{4z_h}
\left(
z_h^4 e^{\chi_h}G_h\Phi_h'^2
+2V_h
\right),
\\
\phi'_h
&=
\frac{
V_{,\phi,h}
-\frac12 z_h^4 e^{\chi_h}
G_{,\phi,h}\Phi_h'^2
}{
z_h^2 f'_h
},
\\
\chi'_h
&=
\frac{z_h}{2}\phi_h'^2
+
\frac{
z_h e^{\chi_h}J_h\Phi_h'^2
}{
2f_h'^2
}.
\end{split}
\end{equation}
These relations provide the regular Taylor data for integrating the solution both outward to the AdS boundary and inward into the black-hole interior.

The radial equations also admit the conserved quantity
\begin{equation}
{\cal Q}
=
\frac{e^{-\chi/2}}{z^2}
\left(
f'-f\chi'
\right)
-
e^{\chi/2}G(\phi)\Phi\Phi',
\qquad
{\cal Q}'=0 .
\label{app:eq:radial_charge}
\end{equation}
Evaluating $\mathcal Q$ at the event horizon and at the AdS boundary gives, respectively, $\mathcal Q=-Ts$ and
$\mathcal Q=3f_3-2\phi_1\phi_2+\mu\rho$. Hence
\begin{equation}
3f_3
-2\phi_1\phi_2
+\mu\rho
=
-Ts .
\label{app:eq:Smarr_general}
\end{equation}
For the source-free solutions considered in this work, $\phi_1=0$, so that
\begin{equation}
3f_3+\mu\rho+Ts=0 .
\label{app:eq:Smarr_sourcefree}
\end{equation}
This identity relates the horizon and boundary data and provides an independent consistency check of the numerical solutions.

\section{Holographic renormalization and black hole thermodynamics}
\label{app:holo_renorm}

For completeness, we summarize the thermodynamic quantities used in the
main text. The renormalized action is obtained by supplementing the bulk
action with the Gibbons--Hawking term and the standard local
counterterms~\cite{deHaro:2000renorm,Skenderis:2002renorm},
\begin{equation}
S_{\rm ren}
=
S_{\rm bulk}
+
2\int_{\partial{\cal M}} d^3x\,\sqrt{-\gamma}\,K
-
\int_{\partial{\cal M}} d^3x\,\sqrt{-\gamma}
\left(
4+{\cal R}[\gamma]+\frac12\phi^2
\right).
\label{app:eq:Sren}
\end{equation}
Here $\gamma_{ab}$ is the induced metric on the regulated boundary,
$K$ is the trace of its extrinsic curvature, and $\mathcal R[\gamma]$ is the corresponding Ricci scalar. For the homogeneous planar solutions
considered in this work, $\mathcal R[\gamma]=0$.

For the source-free branch, holographic renormalization gives
\begin{align}
\mathcal F = f_3,\qquad
\varepsilon = -2f_3,\qquad
p = -f_3,
\label{app:eq:thermo_dictionary}
\end{align}
where $\mathcal F$ is the grand-potential density, $\varepsilon$ the
energy density and $p$ the pressure. These quantities satisfy
\begin{equation}
\varepsilon
=
\mathcal F+Ts+\mu\rho,
\label{app:eq:thermo_identity}
\end{equation}
which is equivalent to the source-free relation \eqref{app:eq:Smarr_sourcefree}. The first law takes the form
\begin{equation}
d\mathcal F
=
-s\,dT-\rho\,d\mu.
\label{app:eq:first_law}
\end{equation}

The normal phase is the planar RN--AdS$_4$ black brane. For a horizon
located at $z=z_h$, its grand-potential density is
\begin{equation}
\mathcal F_{\rm RN}
=
-\frac{1}{z_h^3}
-\frac{\mu^2}{4z_h},
\label{eq:RN_free_energy}
\end{equation}
and its temperature is
\begin{equation}
T_{\rm RN}
=
\frac{3-\mu^2 z_h^2/4}{4\pi z_h}.
\label{eq:RN_temperature}
\end{equation}
At fixed $(T,\mu)$, Eq.~\eqref{eq:RN_temperature} determines the RN
horizon radius and hence the reference grand potential through
Eq.~\eqref{eq:RN_free_energy}.

\section{Local dynamics of the Einstein--Rosen collapse and Josephson regimes}
\label{app:ER_matching}

This appendix reviews the local asymptotic analysis of the
Einstein--Rosen collapse and Josephson regimes following the method of
Ref.~\cite{Hartnoll:2020fhc}, and adapts it to the present setup. These results are
used in Sec.~\ref{sec:interior}.

\subsection{Einstein--Rosen collapse}
\label{subsec:ER_matching}

To resolve the nonlinear region that replaces the RN Cauchy horizon,
we derive the local ER-collapse solution. We consider a narrow layer
around the would-be inner horizon $z=z_\star$, where the scalar remains small but its radial
gradient is enhanced. Since the layer is parametrically thin, explicit
radial factors can be evaluated at $z=z_\star$, while the Maxwell field
is taken to vary slowly, $\Phi\simeq\Phi_o$ and
$\Phi'\simeq E_o e^{-\chi/2}$. At leading order, we use
$J(\phi)\simeq q^2\phi^2$ and retain only the constant part of
$V(\phi)$.

Under these approximations, the scalar equation becomes
\begin{equation}
\left(
f e^{-\chi/2}\phi'
\right)'
+
\frac{q^2\Phi_o^2e^{\chi/2}}{f}\phi
\simeq0.
\label{eq:ER_scalar_equation}
\end{equation}
Introducing $d\eta=e^{\chi/2}dz/f$, the equation becomes
$d^2\phi/d\eta^2+q^2\Phi_o^2\phi=0$. Its solution can be written as
\begin{equation}
\phi
=
\phi_o
\cos\left(
q\Phi_o
\int^z
\frac{e^{\chi/2}}{f}\,dz'
+
\varphi_o
\right),
\label{eq:ER_scalar_solution}
\end{equation}
and the corresponding conserved quantity is
\begin{equation}
e^{-\chi}f^2\phi'^2
+
q^2\Phi_o^2\phi^2
=
q^2\Phi_o^2\phi_o^2.
\label{eq:ER_first_integral}
\end{equation}

To make the metric dynamics explicit, we introduce
\begin{equation}
{\cal Y}
=
e^{-\chi/2},
\qquad
{\cal G}
=
-\frac{f{\cal Y}^2}{z_\star^2},
\qquad
D_o
=
z_\star^4E_o^2-12,
\qquad
k_o
=
\frac{D_o}{4z_\star}.
\label{eq:ER_definitions}
\end{equation}
For the inner-horizon branch of interest, $D_o>0$. Using the scalar
first integral \eqref{eq:ER_first_integral}, the oscillatory scalar
dependence in the Einstein equations is replaced by the constant
$q^2\Phi_o^2\phi_o^2$, and the leading equations reduce to
\begin{equation}
\left(f{\cal Y}\right)'
=
k_o{\cal Y},
\qquad
\chi'
=
\frac{
z_\star q^2\Phi_o^2\phi_o^2
}{
2f^2{\cal Y}^2
}.
\label{app:eq:ER_H_equations}
\end{equation}

It is convenient to introduce the combinations
\begin{equation}
c_1^2
=
\frac{q^2\Phi_o^2\phi_o^2}{D_o},
\qquad
c_2^2
=
\frac{k_o{\cal Y}_m^2}{z_\star^2},
\label{eq:ER_c1c2}
\end{equation}
where ${\cal Y}_m$ is an integration constant. In terms of these
quantities, the reduced Einstein equations integrate to
\begin{equation}
{\cal Y}
=
{\cal Y}_m
\frac{{\cal G}}{{\cal G}+c_1^2},
\qquad
{\cal G}'
=
-c_2^2
\frac{{\cal G}}{{\cal G}+c_1^2}.
\label{app:eq:ER_first_order}
\end{equation}
Integrating the equation for ${\cal G}$ gives
\begin{equation}
{\cal G}
+
c_1^2
\log\left(
\frac{{\cal G}}{c_1^2}
\right)
=
-c_2^2(z-z_o).
\label{eq:ER_normal_form}
\end{equation}
where $z_o$ is an integration constant fixing the radial location of
the collapse layer.

On the outer side of the layer, $\mathcal G \gg c_1^2$, and Eq.~\eqref{app:eq:ER_first_order} reduces to the RN inner-horizon behavior,
\begin{equation}
\mathcal G \simeq c_2^2 (z_o-z),
\label{eq:C10}
\end{equation}
providing the required overlap with the exterior RN geometry.

On the inner side, $\mathcal G \ll c_1^2$, and the same equations yield the exponential collapse
\begin{equation}
\mathcal G \simeq c_1^2
\exp\left[-\frac{z-z_o}{\Delta z_{\rm ER}}\right],
\qquad
\Delta z_{\rm ER}=\frac{c_1^2}{c_2^2}.
\label{eq:ER_width_general}
\end{equation}
Since $-f e^{-\chi}=z_\star^2\mathcal G$, this is precisely the Einstein--Rosen collapse.

\subsection{Quadratic Josephson dynamics}
\label{app:quadratic_Josephson_matching}

On the inner side of the ER collapse, Eq.~\eqref{app:eq:ER_first_order}
gives
\begin{equation}
f e^{-\chi/2}
=
-\frac{\sqrt{D_o z_\star}}{2c_2}
\left(
{\cal G}+c_1^2
\right).
\label{app:eq:ER_H_matching}
\end{equation}
As ${\cal G}/c_1^2\to0$, the combination $f e^{-\chi/2}$ approaches a
finite constant. We therefore write
\begin{equation}
\frac{f e^{-\chi/2}}{z^3}
\simeq
-\frac{1}{c_3},
\qquad
c_3
=
\frac{2c_2}{c_1^2}
\sqrt{\frac{z_\star^5}{D_o}}.
\label{app:eq:c3_matching}
\end{equation}

Taking $\Phi\simeq\Phi_o$ and neglecting the subleading scalar potential,
the scalar equation reduces to
\begin{equation}
\phi''
+
\frac{1}{z}\phi'
+
\frac{q^2\Phi_o^2c_3^2}{z^6}\phi
\simeq0.
\label{app:eq:Josephson_radial_equation}
\end{equation}
Defining $x=|q\Phi_o|c_3/(2z^2)$ brings this equation to the Bessel form.

The general solution is
\begin{equation}
\phi
=
c_4 J_0(x)
+
c_5 Y_0(x).
\label{eq:Bessel_solution}
\end{equation}
For $x\gg1$, the Bessel functions behave as
\begin{align}
J_0(x)
&\simeq
\sqrt{\frac{2}{\pi x}}
\cos\left(x-\frac{\pi}{4}\right),
\qquad
Y_0(x)
\simeq
\sqrt{\frac{2}{\pi x}}
\sin\left(x-\frac{\pi}{4}\right).\label{app:eq:Josephson_solution}
\end{align}
Writing $c_4=A\cos\Theta$ and $c_5=A\sin\Theta$, the large-$x$
solution becomes
\begin{equation}
\phi
\simeq
A
\sqrt{\frac{2}{\pi x}}
\cos\left(
x-\frac{\pi}{4}-\Theta
\right).
\label{app:eq:Josephson_large_x}
\end{equation}

At the outer edge of the oscillatory region, $z\simeq z_\star$, so that
\begin{equation}
x_\star
=
\frac{|q\Phi_o|c_3}{2z_\star^2}
=
\frac{|q\Phi_o|c_2}{c_1^2}
\sqrt{\frac{z_\star}{D_o}}.
\label{app:eq:Josephson_xstar_local}
\end{equation}
Matching the large-$x$ envelope in Eq.~\eqref{app:eq:Josephson_large_x}
to the ER amplitude $\phi_o$ gives
\begin{equation}
A^2
=
\frac{\pi c_2}{2|q\Phi_o|}
\sqrt{D_o z_\star}.
\label{app:eq:Josephson_amplitude_local}
\end{equation}

As the solution evolves inward, $x$ decreases and eventually enters
the small-$x$ regime. Using
\begin{equation}
J_0(x)
=
1+O(x^2),
\qquad
Y_0(x)
=
\frac{2}{\pi}
\left[
\log\left(\frac{x}{2}\right)
+\gamma_E
\right]
+
O(x^2\log x).
\label{app:eq:Bessel_small_x}
\end{equation}
The scalar therefore behaves as
\begin{equation}
\phi
=
c_4
+
\frac{2c_5}{\pi}
\left[
\log\left(\frac{x}{2}\right)
+\gamma_E
\right]
+\cdots .
\label{app:eq:Josephson_small_x}
\end{equation}
Since $x\propto z^{-2}$, the logarithmic branch approaches a constant
radial scalar velocity,
\begin{equation}
z\phi'
\longrightarrow
-\frac{4c_5}{\pi}.
\label{app:eq:Josephson_velocity}
\end{equation}
These asymptotic solutions are used in
Sec.~\ref{subsec:Josephson} to match the ER collapse to the Josephson
regime and to determine the first-Kasner data.

\subsection{Nonlinear phase correction}
\label{app:Josephson_phase}

We next include the leading nonlinear correction from the
St\"uckelberg coupling,
\begin{equation}
J(\phi)
=
q^2\phi^2
+
\eta_4\phi^4
+
O(\phi^6).
\label{eq:Josephson_nonlinear_J}
\end{equation}
Keeping the quartic term in the Josephson regime modifies the Bessel
equation to
\begin{equation}
\phi_{xx}
+
\frac{1}{x}\phi_x
+
\phi
+
\frac{2\eta_4}{q^2}\phi^3
+
\cdots
=
0.
\label{eq:Josephson_nonlinear_equation}
\end{equation}

To isolate the phase evolution, we define $F=\sqrt{x}\phi$, for which
Eq.~\eqref{eq:Josephson_nonlinear_equation} becomes
\begin{equation}
F_{xx}
+
F
+
\frac{2\eta_4}{q^2x}F^3
+
\frac{F}{4x^2}
=
0.
\label{app:eq:Josephson_rescaled_oscillator}
\end{equation}
For $x\gg1$, let $F\simeq a\cos\vartheta$, where $a$ varies only at
subleading order. Retaining the fundamental harmonic of the cubic term
gives
\begin{equation}
\frac{d\vartheta}{dx}
=
1
+
\frac{3\eta_4a^2}{4q^2}\frac{1}{x}
+
O(x^{-2}).
\label{app:eq:Josephson_frequency_shift}
\end{equation}
Hence the quartic interaction generates a logarithmic correction to the
Josephson phase,
\begin{equation}
\vartheta(x)
=
x
+
\frac{3\eta_4a^2}{4q^2}\log x
+
O(1).
\end{equation}
The $O(x^{-2})$ terms contribute only to the finite phase offset.

\section{Zero-temperature infrared solutions}
\label{app:zero_temperature}

For completeness, we derive here the zero-temperature infrared
solutions underlying the low-temperature analysis of
Sec.~\ref{sec:low_temperature}. For the three models considered below,
$G=1$ and $J=9\phi^2$.

It is convenient to introduce
$\varrho=\log z$ and
\begin{equation}
v=z\phi',
\qquad
a=ze^{\chi/2}\Phi,
\qquad
e=z^2e^{\chi/2}\Phi'.
\end{equation}
Denoting derivatives with respect to $\varrho$ by dots, the radial
equations become
\begin{equation}
\begin{split}
\dot\chi
&=
\frac{v^2}{2}
+
\frac{Ja^2}{2f^2},
\qquad
\dot\phi
=
v,
\\
\dot v
&=
-\frac{e^2+2V}{4f}v
+
\frac{V_{,\phi}}{f}
-
\frac{J_{,\phi}a^2}{2f^2},
\\
\dot f
&=
3f
+
\frac{f\dot\chi}{2}
+
\frac{e^2+2V}{4},
\\
\dot a
&=
\left(
1+\frac{\dot\chi}{2}
\right)a+e,
\qquad
\dot e
=
2e+\frac{Ja}{f}.
\end{split}
\label{app:eq:IR_autonomous}
\end{equation}
The source-free zero-temperature solution also satisfies
\begin{equation}
(12-v^2)f^2
+
(e^2+2V-4ae)f
-
Ja^2
=
0.
\label{app:eq:IR_zero_charge}
\end{equation}

The three models approach a cohesive AdS$_4$ fixed point, a charged
Lifshitz fixed point, and a running-scalar solution, respectively.

\subsection{Cohesive AdS$_4$ infrared solution}
\label{app:AdS4_endpoint}

For the cohesive endpoint, the electric sector is irrelevant and the
scalar approaches a stationary point of the potential. Setting
$v=a=e=0$ in Eq.~\eqref{app:eq:IR_autonomous} gives
\begin{equation}
V_{,\phi}(\phi_*)=0,
\qquad
f_*=-\frac{V_*}{6},
\qquad
L_{\rm IR}^2=-\frac{6}{V_*}.
\label{app:eq:AdS4_fixed_point}
\end{equation}
For $\mathcal M_{\rm AdS}$,
\begin{equation}
\phi_*=1,
\qquad
f_*=\frac{10}{9},
\qquad
L_{\rm IR}^2=\frac{9}{10}.
\end{equation}

Linear scalar and vector perturbations behave as
$\delta\phi\propto z^{-\delta_\phi}$ and
$\Phi\propto z^{-\delta_A}$, with
\begin{equation}
\delta_\phi(\delta_\phi+3)
=
\frac{V_{,\phi\phi,*}}{f_*},
\qquad
\delta_A(\delta_A+1)
=
\frac{J_*}{f_*}.
\label{app:eq:AdS4_irrelevant_exponents}
\end{equation}
For the present model,
\begin{equation}
\delta_\phi=1.5741,
\qquad
\delta_A=2.3896.
\label{app:eq:AdS4_irrelevant_values}
\end{equation}
Since $\delta_A>\delta_\phi$, the correction sourced by the vector mode
decays faster than the leading scalar deformation. The approach to the
IR fixed point is therefore controlled by $\delta_\phi$.

To determine the global zero-temperature solution, we shoot from the IR
fixed point toward the AdS boundary. The IR expansion is specified by
the leading irrelevant scalar deformation,
\begin{equation}
\phi
=
1-A_\phi z^{-\delta_\phi}
+\cdots,
\qquad
\chi
=
\chi_{\rm IR}
+\cdots,
\label{app:eq:adsirphi}
\end{equation}
while the UV solution is required to satisfy the source-free boundary
condition together with the normalization $\chi(0)=0$.

The shooting parameter is adjusted so that the non-normalizable scalar
mode vanishes at the AdS boundary. The resulting zero-temperature
domain wall gives
\begin{equation}
A_\phi=1.0020,
\qquad
\chi_{\rm IR}=0.35667.
\label{app:eq:AdS4_global_data}
\end{equation}

\subsection{Finite-scalar Lifshitz infrared solution}
\label{app:Lifshitz_endpoint}

A Lifshitz infrared fixed point is invariant under
$t\to\lambda^{\mathfrak z}t$, $x^i\to\lambda x^i$, and
$z\to\lambda z$. Comparing this scaling with our metric and gauge-field
ansatz fixes the radial dependence of the fields to
\begin{equation}
\phi=\phi_L,
\qquad
f=f_L,
\qquad
\chi
=
2(\mathfrak z-1)\log z+\chi_L,
\qquad
\Phi
=
\Phi_L z^{-\mathfrak z}.
\label{app:eq:Lifshitz_ansatz}
\end{equation}
Substitution into \eqref{app:eq:IR_autonomous} gives
\begin{equation}
\begin{split}
J_L
&=
2\mathfrak z f_L,
\qquad
f_L
=
-\frac{V_L}{\mathfrak z^2+\mathfrak z+4},
\\
e^{\chi_L}\Phi_L^2
&=
\frac{2(\mathfrak z-1)f_L}{\mathfrak z},
\qquad
V_{,\phi,L}
=
\frac{\mathfrak z-1}{\mathfrak z}
J_{,\phi,L}.
\end{split}
\label{app:eq:Lifshitz_algebraic}
\end{equation}
For $\mathcal M_{\rm Lif}$,
\begin{equation}
\phi_L=\frac{3}{\sqrt{35}},
\qquad
f_L=\frac{81}{140},
\qquad
\mathfrak z=2.
\end{equation}

To determine how the zero-temperature solution approaches the Lifshitz
fixed point, we perturb the autonomous variables as
\begin{equation}
Y(\varrho)
=
Y_L
+
\delta Y(\varrho),
\qquad
Y\equiv(\phi,v,f,a,e),
\end{equation}
where $Y_L$ denotes the Lifshitz fixed point. Linearizing
$\dot Y=F(Y)$ around $Y_L$ gives
\begin{equation}
\delta\dot Y
=
M_L\,\delta Y,
\qquad
M_L
\equiv
\left.
\frac{\partial F}{\partial Y}
\right|_{Y_L}.
\end{equation}
Looking for power-law perturbations
$\delta Y=k_\sigma z^\sigma=e^{\sigma\varrho}k_\sigma$, one obtains the
eigenvalue problem
\begin{equation}
M_L k_\sigma
=
\sigma k_\sigma.
\label{app:eq:Lifshitz_linearization}
\end{equation}
The eigenvalue problem yields
\begin{equation}
\sigma
=
\left\{
-6.80397906,\,
-0.55721243,\,
4,\,
4.55721243,\,
10.80397906
\right\}.
\label{app:eq:Lifshitz_spectrum}
\end{equation}
Since the IR corresponds to $z\to\infty$, the negative eigenvalues describe
irrelevant deformations. The slowest approach to the Lifshitz fixed point is
therefore governed by
\begin{equation}
\lambda_1
=
0.55721243.
\end{equation}

The local analysis fixes the irrelevant eigendirection but not its amplitude.
We normalize the corresponding eigenvector by
$(k_{-\lambda_1})_\phi=1$ and shoot the full zero-temperature solution toward
the AdS boundary, imposing the source-free UV condition and $\chi(0)=0$.
The resulting domain wall approaches the Lifshitz fixed point as
\begin{equation}
Y_0-Y_L
=
A_s k_{-\lambda_1}z^{-\lambda_1}
+\cdots,
\qquad
A_s=-0.38847808,
\label{app:eq:Lifshitz_global_mode}
\end{equation}
with
\begin{equation}
\chi_0-2\log z
\longrightarrow
\chi_L,
\qquad
\chi_L=-8.1730469.
\label{app:eq:Lifshitz_global_matching}
\end{equation}

\subsection{Running-scalar infrared solution}
\label{app:running_endpoint}

For $\mathcal M_{\rm run}$, the scalar does not approach a finite IR
fixed point. Instead $\phi\to+\infty$, while the electric field becomes
subleading. Setting $a=e=0$ in
Eq.~\eqref{app:eq:IR_zero_charge} gives
\begin{equation}
f
=
\frac{\phi^2+6}{2(3-v^2/4)},
\qquad
v\frac{dv}{d\phi}
=
\left(
3-\frac{v^2}{4}
\right)
\left(
v-\frac{4\phi}{\phi^2+6}
\right).
\label{app:eq:running_neutral_reduction}
\end{equation}
The same equations imply
$d\log f/d\varrho=\dot\chi=v^2/2$, and hence
\begin{equation}
fe^{-\chi}
=
c_{\rm IR}^2,
\end{equation}
where $c_{\rm IR}$ is fixed by the complete zero-temperature solution.

Solving Eq.~\eqref{app:eq:running_neutral_reduction} at large $\phi$
gives
\begin{equation}
v
=
\frac4\phi
-
\frac{88}{3\phi^3}
+
\frac{880}{3\phi^5}
+
O(\phi^{-7}),
\qquad
f
=
\frac{\phi^2}{6}
+
\frac{11}{9}
+
O(\phi^{-2}).
\label{app:eq:running_velocity_series}
\end{equation}
Since $v=d\phi/d\varrho$, integration yields
\begin{equation}
\phi^2
=
8\varrho
-
\frac{22}{3}\log\varrho
+
C_\phi
+
O\!\left(
\frac{\log\varrho}{\varrho}
\right).
\label{app:eq:running_scalar_expansion}
\end{equation}

It remains to verify that the electric sector is indeed subleading.
Linearizing the Maxwell equation on this neutral background gives
\begin{equation}
\ddot\Phi
+
\left(
\frac{\dot\chi}{2}-1
\right)\dot\Phi
-
\frac{q^2\phi^2}{f}\Phi
=
0.
\label{app:eq:running_Maxwell}
\end{equation}
For
$\Phi\sim e^{-\beta\varrho}\varrho^\gamma$, the asymptotic indices obey
\begin{equation}
\beta(\beta+1)=6q^2,
\qquad
\gamma
=
\frac{11q^2-\beta}{2(2\beta+1)}.
\label{app:eq:running_Maxwell_indices}
\end{equation}
For $q=3$,
\begin{equation}
\beta=6.86546,
\qquad
\gamma=3.12725.
\end{equation}
On the neutral running background, $f\sim\varrho$ and
$e^{\chi/2}\sim\varrho^{1/2}$. Hence the Maxwell mode
$\Phi\sim z^{-\beta}\varrho^\gamma$ gives
$a,\ e
\sim
z^{1-\beta}\varrho^{\gamma+1/2}$.
Since $\beta>1$, both electric variables vanish as $z\to\infty$, and
their quadratic contributions to the Einstein equations are
parametrically smaller than the logarithmically varying neutral
background. The electric sector is therefore subleading in the IR.

The local IR expansion determines the asymptotic form but not the
constants $c_{\rm IR}$ and $C_\phi$. They are fixed by the complete
source-free zero-temperature domain wall, with
$c_{\rm IR}^2=\lim_{z\to\infty}f e^{-\chi}$ and $C_\phi$ extracted from
Eq.~\eqref{app:eq:running_scalar_expansion}. Imposing the UV
normalization $\chi(0)=0$, we find
\begin{equation}
c_{\rm IR}^2
=
0.81612932,
\qquad
C_\phi
=
-16.6830927.
\label{app:eq:running_global_data}
\end{equation}


\end{document}